\documentclass[traditabstract,printers]{aa}

\usepackage{txfonts}

\usepackage{natbib}
\usepackage{graphicx}

\usepackage{aalongtable}

\begin{document}
\title{SuperWASP observations of pulsating Am stars\thanks{
An extended version of Table~\ref{table:Pulsating Am stars} containing all
the detected frequencies and amplitudes is only available in electronic form
at the CDS via anonymous ftp to cdsarc.u-strasbg.fr (130.79.128.5)
or via http://cdsweb.u-strasbg.fr/cgi-bin/qcat?J/A+A/}}
\author
{B.~Smalley\inst{1},
D.~W.~Kurtz\inst{2},
A.~M.~S.~Smith\inst{1},
L.~Fossati\inst{3},
D.~R.~Anderson\inst{1},
S.~C.~C.~Barros\inst{4},
O.~W.~Butters\inst{5},
A.~Collier Cameron\inst{6},
D.~J.~Christian\inst{7},
B.~Enoch\inst{6},
F.~Faedi\inst{4},
C.~A.~Haswell\inst{3},
C.~Hellier\inst{1},
S.~Holmes\inst{3},
K.~Horne\inst{6},
S.~R.~Kane\inst{8},
T.~A.~Lister\inst{9},
P.~F.~L.~Maxted\inst{1},
A.~J.~Norton\inst{3},
N.~Parley\inst{6},
D.~Pollacco\inst{4},
E.~K.~Simpson\inst{4},
I.~Skillen\inst{10},
J.~Southworth\inst{1},
R.~A.~Street\inst{9},
R.~G.~West\inst{5},
P.~J.~Wheatley\inst{11},
P.~L.~Wood\inst{1}}

\authorrunning{B. Smalley et al.}
\titlerunning{SuperWASP observations of pulsating Am stars}


\institute{
Astrophysics Group, Keele University, Staffordshire, ST5\,5BG, United Kingdom
\and
Jeremiah Horrocks Institute of Astrophysics, University of Central Lancashire,
Preston PR1\,2HE, UK
\and
Department of Physics \& Astronomy, The Open University, Walton Hall, Milton
Keynes, MK7 6AA, UK
\and
Astrophysics Research Centre, Main Physics Building, School of Mathematics \&
Physics, Queen's University, University Road, Belfast, BT7 1NN, UK
\and
Department of Physics \& Astronomy, University of Leicester, Leicester, LE1
7RH, UK
\and
SUPA, School of Physics \& Astronomy, University of St. Andrews,
North Haugh, Fife, KY16 9SS, UK
\and
Department of Physics \& Astronomy, California State University, Northridge,
CA, 91330, USA
\and
NASA Exoplanet Science Institute, Caltech, MS 100-22, 770 South Wilson Avenue,
Pasadena, CA, 91125, USA
\and
Las Cumbres Observatory Global Telescope Network, 6740 Cortona Drive, Suite 102,
Goleta, CA, 93117, USA
\and
Isaac Newton Group of Telescopes, Apartado de Correos 321, 38700 Santa Cruz de
la Palma, Tenerife, Spain
\and
Department of Physics, University of Warwick, Coventry, CV4 7AL, UK}

\date{Received date / accepted date}

\abstract{We have studied over 1600 Am stars at a photometric precision of
1\,mmag with SuperWASP photometric data. Contrary to previous belief, we find
that around 200 Am stars are pulsating $\delta$\,Sct and $\gamma$\,Dor
stars, with low amplitudes that have been missed in previous, less extensive
studies. While the amplitudes are generally low, the presence of pulsation
in Am stars places a strong constraint on atmospheric convection, and may
require the pulsation to be laminar. While some pulsating Am stars have been
previously found to be $\delta$\,Sct stars, the vast majority of Am stars
known to pulsate are presented in this paper. They will form the basis of
future statistical studies of pulsation in the presence of atomic diffusion.}


\maketitle

\section{Introduction}

In the region of the Hertzsprung-Russell (HR)~diagram where the Cepheid instability strip extends
across the main sequence, there is a complex relationship between stellar
pulsation and atmospheric abundance anomalies that is not fully understood. This
region ranges from the early A stars to mid-F stars in spectral type, and from
the zero age main sequence to the terminal age main sequence in luminosity.
Found here are the strongly magnetic chemically peculiar Ap and Fp stars, the
non-magnetic metallic-lined Am stars, the rarer metal-deficient $\lambda$\,Boo
stars, the pulsating $\delta$\,Sct stars, $\gamma$\,Dor stars and rapidly
oscillating Ap (roAp) stars. Much has been written about these stars and their
physics, which we briefly summarise here. For more detailed discussions see
\citet{2006A&A...455..303J}, \citet{2000BaltA...9..253K} and
\cite{1989MNRAS.238.1077K,1978ApJ...221..869K,1976ApJS...32..651K}.

Most stars in the main-sequence region of the instability strip are normal
abundance $\delta$\,Sct stars with relatively high rotational velocities --
usually $v \sin i \ge 100$\,km\,s$^{-1}$. A large fraction of A stars are Am
stars, peaking at around 50 per cent at A8, but Am stars are believed either not
to pulsate as $\delta$\,Sct stars, or may do so with much smaller amplitudes
than the normal abundance $\delta$\,Sct stars. Am stars are mostly found in
short period binary systems with orbital periods between $1-10$\,d, causing
synchronous rotation with $v \sin i \le 120$\,km\,s$^{-1}$
\citep{2009AJ....138...28A}; a few single Am stars with similar slow rotation
are known.

The magnetic Ap stars are rarer, constituting less than 10 per cent of the A
stars. They have very strong global magnetic fields and are often roAp stars
with high overtone p\,mode pulsations with much shorter periods than the
$\delta$\,Sct stars. No Ap star is known to be a $\delta$\,Sct star. Our
physical understanding is that atomic diffusion -- radiative levitation and
gravitational settling -- stabilises the slowly rotating Am and Ap stars so that
low overtone p\,modes are not excited; particularly important in this context is
the gravitational settling of helium from the He\,\textsc{ii} ionisation zone
where the $\kappa$-mechanism drives the pulsation of $\delta$\,Sct stars (see
\citealt{2010aste.book.....A}). Otherwise, the more rapidly rotating stars
remain mixed because of turbulence induced by meridional circulation and are
excited by the $\kappa$-mechanism \citep{2000A&A...360..603T}.

The understanding of the relationship of the long-established $\delta$\,Sct
stars to the more recently discovered $\gamma$\,Dor stars is currently in flux.
Previously, the $\delta$\,Sct stars were known as p\,mode pulsators, while the
$\gamma$\,Dor stars were known as g\,mode pulsators. The instability strips for
these classes of stars partially overlap, and some ``hybrid'' stars were
discovered with pulsation in both p\,modes and g\,modes. A striking case is that
of HD\,8801, which is an Am star that shows both $\delta$\,Sct and $\gamma$\,Dor
p-mode and g-mode pulsation \citep{2005AJ....129.2026H}.

Hybrid stars that show both p\,modes and g\,modes are of particular interest
asteroseismically because the p\,modes characterise the conditions primarily in
the outer part of the star, while the g\,modes test the core conditions. Now
with data from the {\it Kepler} Mission, which is obtaining nearly continuous
data for over 150\,000 stars for 3.5\,y, mostly with 30-min cadence, but for 512
stars with 1-min cadence \citep{2010PASP..122..131G}, the Kepler Asteroseismic
Science Consortium (KASC) is studying numbers of $\delta$\,Sct stars and
$\gamma$\,Dor stars at $\mu$mag precision. It is becoming clear that hybrid
stars are common and may be the norm, so that the classes of $\delta$\,Sct and
$\gamma$\,Dor stars are merging \citep{2010ApJ...713L.192G}. Interestingly, the
latter authors find a possible correlation among the hybrid stars and Am
spectral classification.

The {\it Kepler} Mission through KASC will model individual Am stars that are
$\delta$\,Sct pulsators with data of such high precision that new insight into
the physics of the relationship between atomic diffusion and p\,mode pulsation
will be obtained. But {\it Kepler} has a limited number of Am stars in its
105\,deg$^2$ field-of-view. Another complementary source of information is to
look at the statistics of pulsation in Am stars over the entire sky. That is now
possible with the highly successful SuperWASP planetary transit-finding
programme \citep{2006PASP..118.1407P} that has surveyed a large fraction of both
the northern and southern skies. There now exists in the SuperWASP archive over
290 billion photometric measurements for more than 30 million stars. These light
curves encompass many types of stars, including the A stars in general, and Am
stars in particular.

In this paper we have selected Am stars from the \citet{2009A&A...498..961R}
catalogue of peculiar stars for which we have at least 1000 data points in
SuperWASP light curves. While we do not detect pulsation in all of our programme
stars, for around 200 metallic-lined stars out of over 1600 tested we find
$\delta$\,Sct pulsation. This is contrary to previous understanding that Am
stars are constant in brightness. The reason we have gained this new
understanding is that there has been no previous survey of so many Am stars, and
previous studies have not all reached the SuperWASP detection threshold of only
1\,mmag.

Many Am stars therefore {\it do} pulsate, generally with lower amplitude than
normal abundance $\delta$\,Sct stars. This amplitude difference is still to be
understood in terms of atomic diffusion reducing pulsation driving for the
slowly rotating Am stars, but there is not a complete lack of pulsation.
That, has implications for turbulence in the diffusive layers and may
require that the pulsation be laminar. Some striking examples of metallic-lined
stars with relatively high pulsation amplitude (these are rare) address this
question further, such as HD\,188136 (\citealt{1980MNRAS.193...29K};
\citealt{1981ApJ...247..969W}) and HD\,40765 \citep{1995MNRAS.276..199K}. More
constraints on the physics of the interaction of pulsation and atomic diffusion
may also be found in stars that show {\it no} $\delta$\,Sct p\,modes or
$\gamma$\,Dor g\,modes at precisions of $\mu$mag. Some such A stars are known in
the CoRoT and {\it Kepler} data sets, but in-depth studies have not yet been
made, hence discussions of these have yet to be published.

The combination of the all-sky mmag precision of SuperWASP with the $\mu$mag
precision of CoRoT and {\it Kepler} on selected stars, calls for new attempts to
model the physics of the interaction of pulsation, rotation and atomic diffusion
in the A stars.

\section{Observations}

The WASP project is surveying the sky for transiting extrasolar planets
\citep{2006PASP..118.1407P} using two robotic telescopes, one at the
Observatorio del Roque de los Muchachos on the island of La Palma in the Canary
Islands, and the other at the Sutherland Station, South African Astronomical
Observatory (SAAO). Both telescopes consist of an array of eight 200-mm, f/1.8
Canon telephoto lenses and Andor CCDs, giving a field of view of $7.8\degr
\times 7.8\degr$ and pixel size of around 14\,\arcsec. The observing strategy is
such that each field is observed with a typical cadence of the order of 10\,min.
WASP provides good quality photometry with a precision exceeding 1~per~cent per
observation in the approximate magnitude range $9 \le V \le 12$.

The SuperWASP data reduction pipeline is described in detail in
\cite{2006PASP..118.1407P}. The aperture-extracted photometry from each camera
on each night are corrected for primary and secondary extinction, instrumental
colour response and system zero-point relative to a network of local secondary
standards. The resultant pseudo-$V$ magnitudes are comparable to Tycho $V$
magnitudes. Additional systematic errors affecting all the stars are identified
and removed using the SysRem algorithm of \cite{2005MNRAS.356.1466T}. The final
light curves are stored in the WASP project's searchable archive
\citep{2010A&A...520L..10B}.

\section{Am star selection and analysis}

We have selected Am stars from the \citet{2009A&A...498..961R} catalogue of
peculiar stars for which we have data in the WASP archive and when individual
light curves have at least 1000 data points (i.e. for a single camera and during
a single season). Any stars known, or found, to be eclipsing binary systems were
excluded from the analysis. Stars were also rejected when two approximately
equal brightness stars were within the 3.5-pixel ($\sim$50\,\arcsec) SuperWASP
photometry aperture. However, unresolved close pairs in DSS images (separation
$\la$2\,\arcsec) and systems with fainter companions ($\ga2$\,mag) were retained.

For each individual light curve, periodograms were calculated using the fast
computation of the Lomb periodogram method of \cite{1989ApJ...338..277P} as
implemented in the Numerical Recipes {\sc fasper} routine
\citep{1992nrfa.book.....P}. Spectral window functions were also calculated,
in order to identify peaks which had arisen due to the gaps in the
observations. The periodograms were examined for any evidence of variability.
Stars were rejected if the false alarm probability of the strongest peaks
exceeded 0.1 \citep{1986ApJ...302..757H}. The remaining stars were examined in
more detail using the {\sc Period04} program \citep{2005CoAst.146...53L}.
For stars in which variability was confirmed, frequencies continued
to be selected so long as their amplitude was $>4$ times the average background
of the pre-whitened residuals \citep{1993A&A...271..482B}. Formal
uncertainties on frequencies and amplitudes were obtained from the least-squares
fitting using the method of \cite{1999DSSN...13...28M}.

Of the 1620 Am stars initially selected, a total of 227 (14\% of the total) have
been found to pulsate. The remaining 1393 stars were deemed as ``not found to
pulsate'', since low-level pulsation could be present below the SuperWASP
detection limits. Table~\ref{table:Pulsating Am stars} provides a summary of the
pulsating Am stars. The individual
periodograms and phase-folded lightcurves are presented in
Fig.~1. 

\addtocounter{table}{1}
\addtocounter{figure}{1}

\section{Stellar parameters}
\label{parameters}

To place stars on the HR~diagram we require values of $T_\mathrm{eff}$
and $\log L$. For stars with $uvby\beta$ photometry in the
\cite{1998A&AS..129..431H} catalogue, we used the {\sc uvbybeta} code of
\citet{1985CommULO..78} to obtain de-reddened indices, and the ($b-y$, $c_0$)
grids of \cite{1997A&A...328..349S} to determine $T_\mathrm{eff}$ and $\log g$.
For stars with only $uvby$ photometry the above procedure was used but without
the de-reddening step. For stars without $uvby$ photometry, Geveva photometry
from \cite{1999yCat.2169....0R} was used and the calibration of
\cite{1997A&AS..122...51K} used to determine $T_\mathrm{eff}$ and $\log g$,
assuming zero reddening. In all of the above cases, the
\cite{2010A&ARv..18...67T} relations were used to determine $\log L$. For stars
without suitable intermediate-band photometry, but with Hipparcos parallaxes
\citep{2007A&A...474..653V}, spectral energy distributions (SEDs) were
constructed using literature broad-band photometry. Values of $T_\mathrm{eff}$
were determined by fitting Kurucz flux distributions to the SEDs and $\log L$
determined from the bolometric flux at the earth ($f_\oplus$) and the Hipparcos
parallax. The typical uncertainties are estimated to be $\pm 200$\,K in
$T_\mathrm{eff}$ ($\pm 0.01$ in $\log T_\mathrm{eff}$) and $\pm 0.25$ in $\log
L$. The stellar parameters are given in Table~\ref{table:Pulsating Am stars}. In
total around a third of the Am stars investigated have stellar parameters
determined.

\section{Am stars in Kepler field}

The sky coverage of the SuperWASP survey overlaps with a large fraction of the
{\it Kepler} field. For Am stars with light curves in both the {\it Kepler}
Public archive and the SuperWASP database we have compared the frequencies and
amplitudes. This allows us to evaluate the detection limits of SuperWASP. Of the
10 stars with both {\it Kepler} and SuperWASP data, four have clear pulsations
with amplitudes $\ga 1$\,mmag (Table~\ref{WASP-Kepler-Stars}), while the other
six stars have amplitudes below the SuperWASP detectability limit.

\begin{table}

\caption{Am stars with both SuperWASP and {\it Kepler} data.}

\centering
\begin{tabular}{cccc} \hline \hline
KIC     & Ren ID & Max Amp & Ref     \\
        &        & (mmag)  &         \\\hline
 9204718& 49340  & 0.13    & Bal     \\
11445913& 49650  & 2.5     & Cat,Bal \\
 9272082& 49840  & $<$0.01 & Bal     \\
12253106& 50070  & $<$0.01 &         \\
 9764965& 50230  & 1.0     &         \\
 8881697& 50420  & 1.9     &         \\
11402951& 50670  & 1.2     & Cat,Bal \\
 9349245& 51233  & $<$0.1  &         \\
 8703413& 51640  & $<$0.1  & Bal     \\
 8323104& 52260  & $<$0.1  & Bal     \\\hline
\end{tabular}
\tablefoot{The second column gives the identification number (Ren ID) from the
\citet{2009A&A...498..961R} catalogue. Column 3 gives the amplitude (Max Amp) of
the highest peak in the {\it Kepler} periodogram. Column 4 gives reference to
published {\it Kepler} data: Cat: \citet{2011MNRAS.411.1167C}, Bal:
\citet{2011MNRAS.414..792B}}
\label{WASP-Kepler-Stars}
\end{table}

\begin{table}

\caption{Comparison between frequencies and amplitudes found in the {\it Kepler}
and SuperWASP data for the four Am stars common to both.}

\centering
\begin{tabular}{ccccc} \hline \hline
&\multicolumn{2}{c}{{\it Kepler}}&\multicolumn{2}{c}{SuperWASP}\\\hline
& Freq. & Amp.\tablefootmark{a} & Freq. & Amp. \\
&(d$^{-1}$) & (mmag) & (d$^{-1}$) & (mmag) \\\hline
\multicolumn{5}{c}{Ren ID 49650 (KIC 11445913, 1SWASP J190540.61+491820.7)}\\\hline
$f_{ 1}$& 31.5577 $\pm$ 0.0003 & 2.8 & 31.5577 $\pm$ 0.0001 & 3.2 $\pm$ 0.1 \\
$f_{ 2}$& 25.3799 $\pm$ 0.0007 & 1.1 & 25.3769 $\pm$ 0.0001 & 1.2 $\pm$ 0.1 \\
$f_{ 3}$& 22.1307 $\pm$ 0.0009 & 0.8 & 22.1306 $\pm$ 0.0002 & 1.0 $\pm$ 0.1 \\
$f_{ 4}$& 37.8182 $\pm$ 0.0011 & 0.6 & \\
$f_{ 5}$& 29.7394 $\pm$ 0.0012 & 0.6 & \\
\hline
\multicolumn{5}{c}{Ren ID 50230 (KIC 9764965, 1SWASP J191724.91+463535.2)}\\\hline
$f_{ 1}$& 27.1777 $\pm$ 0.0001 & 1.1 & 27.1778 $\pm$ 0.0001 & 1.2 $\pm$ 0.1 \\
$f_{ 2}$& 21.3819 $\pm$ 0.0002 & 0.6 & 22.3891 $\pm$ 0.0001 & 0.9 $\pm$ 0.1 \\
$f_{ 3}$& 31.9895 $\pm$ 0.0002 & 0.4 & 31.9902 $\pm$ 0.0001 & 0.9 $\pm$ 0.1 \\
$f_{ 4}$& 19.9579 $\pm$ 0.0004 & 0.2 \\
\hline
\multicolumn{5}{c}{Ren ID 50420 (KIC 8881697, 1SWASP J192136.03+450706.8)}\\
\hline
$f_{ 1}$& 16.5567 $\pm$ 0.0003 &  1.9 & 16.5565 $\pm$ 0.0005 & 2.1 $\pm$ 0.1 \\
$f_{ 2}$& 32.0477 $\pm$ 0.0004 &  1.5 & 32.0481 $\pm$ 0.0006 & 1.6 $\pm$ 0.1 \\
$f_{ 3}$& 25.2105 $\pm$ 0.0005 &  1.2 & 25.2064 $\pm$ 0.0008 & 1.2 $\pm$ 0.1 \\
$f_{ 4}$& 30.0120 $\pm$ 0.0006 &  1.1 & 30.0111 $\pm$ 0.0009 & 1.1 $\pm$ 0.1 \\
$f_{ 5}$& 34.3647 $\pm$ 0.0007 &  0.9 & 34.3661 $\pm$ 0.0009 & 1.0 $\pm$ 0.1 \\
$f_{ 6}$& 30.6537 $\pm$ 0.0008 &  0.9 & 30.6569 $\pm$ 0.0010 & 0.9 $\pm$ 0.1 \\
$f_{ 7}$& 28.8044 $\pm$ 0.0009 &  0.7 & 27.8049 $\pm$ 0.0011 & 0.8 $\pm$ 0.1 \\
$f_{ 8}$& 34.0106 $\pm$ 0.0009 &  0.7 &  \\
$f_{ 9}$& 27.4073 $\pm$ 0.0010 &  0.7 &  \\
$f_{10}$& 16.0119 $\pm$ 0.0013 &  0.5 &  \\
\hline
\multicolumn{5}{c}{Ren ID 50670 (KIC 11402951, 1SWASP J192732.81+491523.5)}\\\hline
$f_{ 1}$& 23.8493 $\pm$ 0.0004 & 1.3 & 23.8464 $\pm$ 0.0008 & 1.4 $\pm$ 0.2 \\
$f_{ 2}$& 23.2770 $\pm$ 0.0004 & 1.1 & 23.2790 $\pm$ 0.0008 & 1.4 $\pm$ 0.2 \\
$f_{ 3}$& 27.4616 $\pm$ 0.0007 & 0.7 & 27.4643 $\pm$ 0.0012 & 1.0 $\pm$ 0.2 \\
$f_{ 4}$& 15.1001 $\pm$ 0.0007 & 0.7 \\
$f_{ 4}$& 14.4967 $\pm$ 0.0009 & 0.5 \\
\hline
\end{tabular}
\tablefoot{
\tablefoottext{a}{Uncertainties on {\it Kepler} Amplitudes are all
$<0.05$\,mmag.}
}
\label{WASP-Kepler-analysis}
\end{table}

\begin{figure*}
\centering
\includegraphics[height=7cm,width=4.5cm,angle=-90]{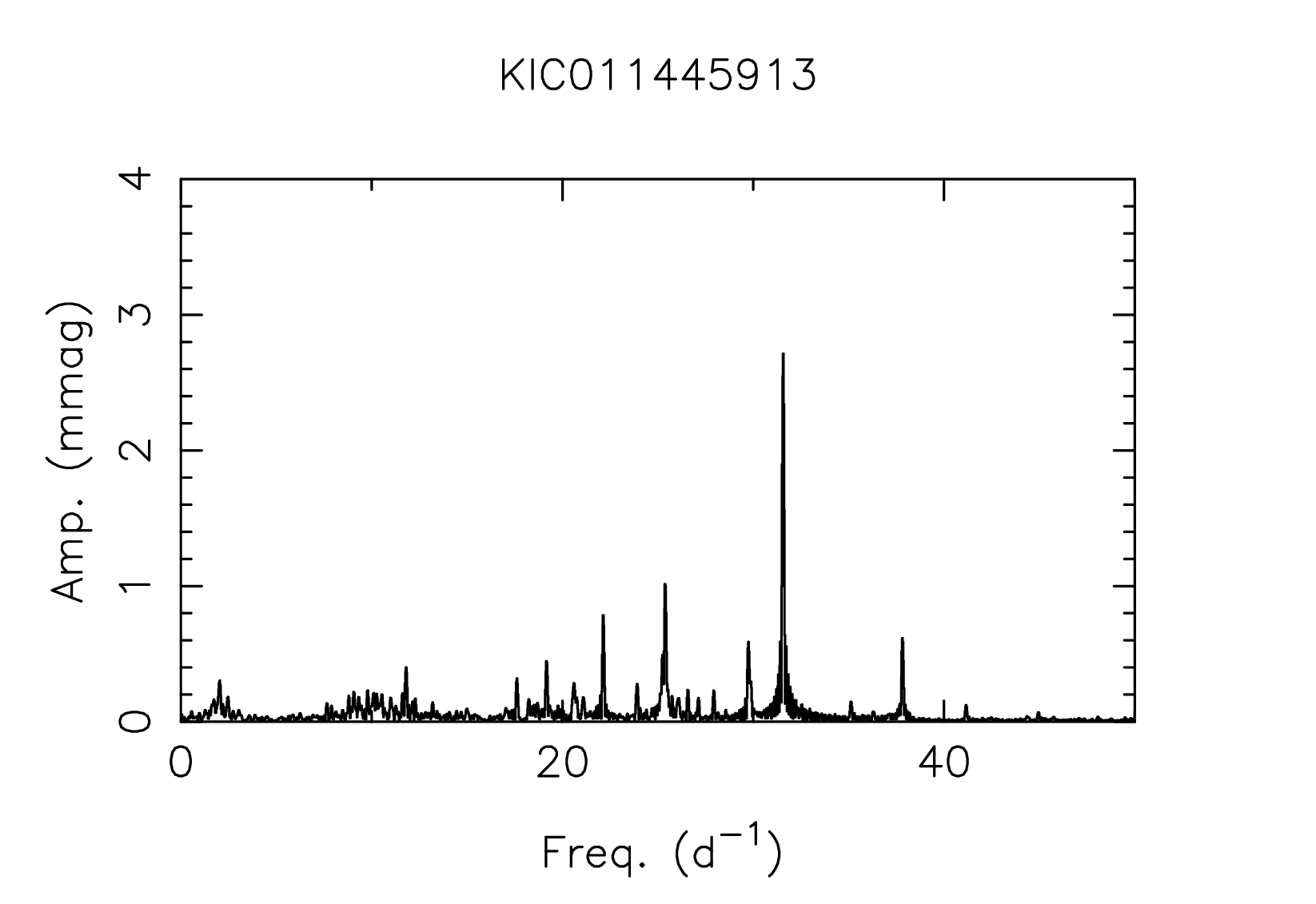}
\includegraphics[height=7cm,width=4.5cm,angle=-90]{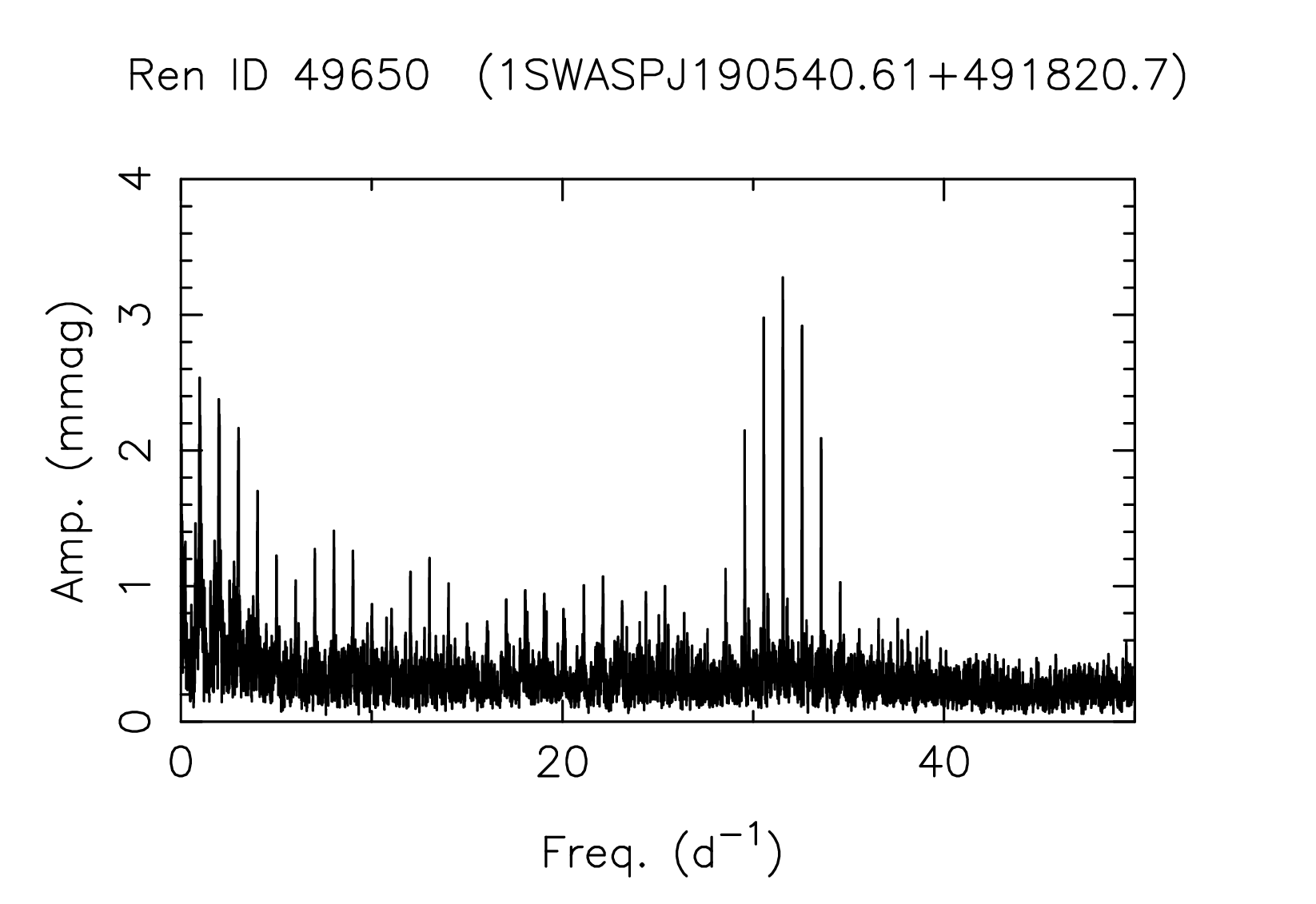}\\
\includegraphics[height=7cm,width=4.5cm,angle=-90]{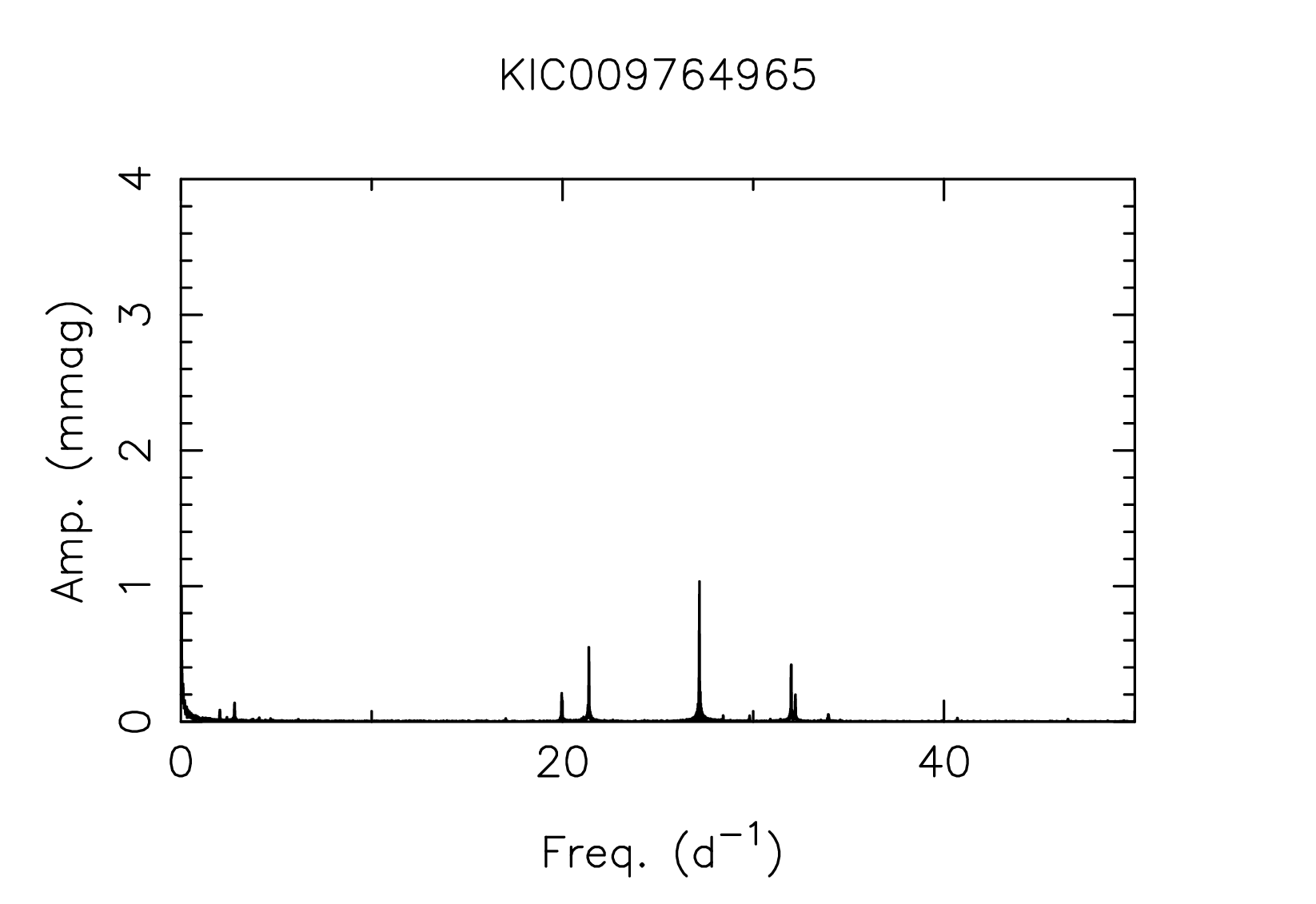}
\includegraphics[height=7cm,width=4.5cm,angle=-90]{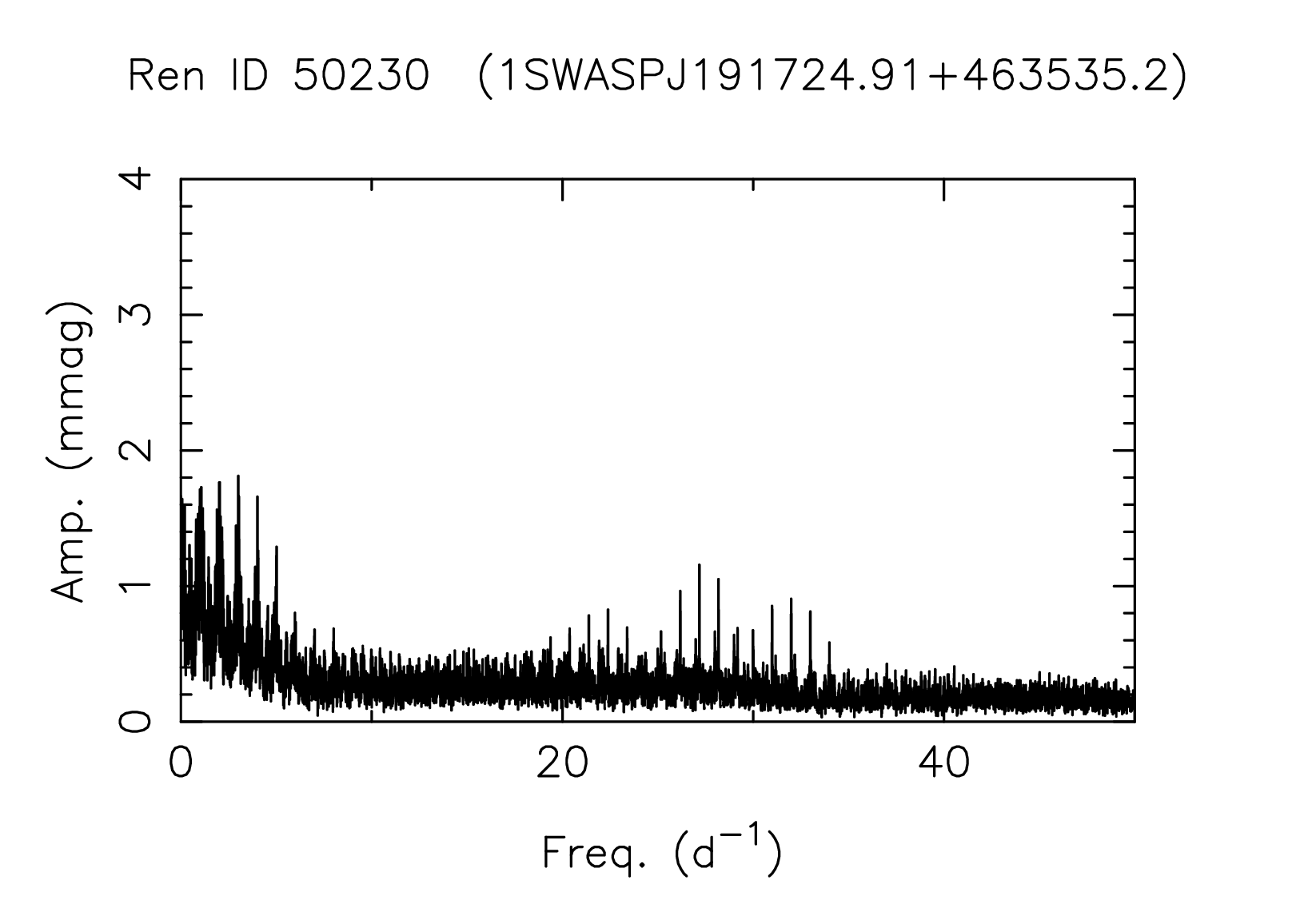}\\
\includegraphics[height=7cm,width=4.5cm,angle=-90]{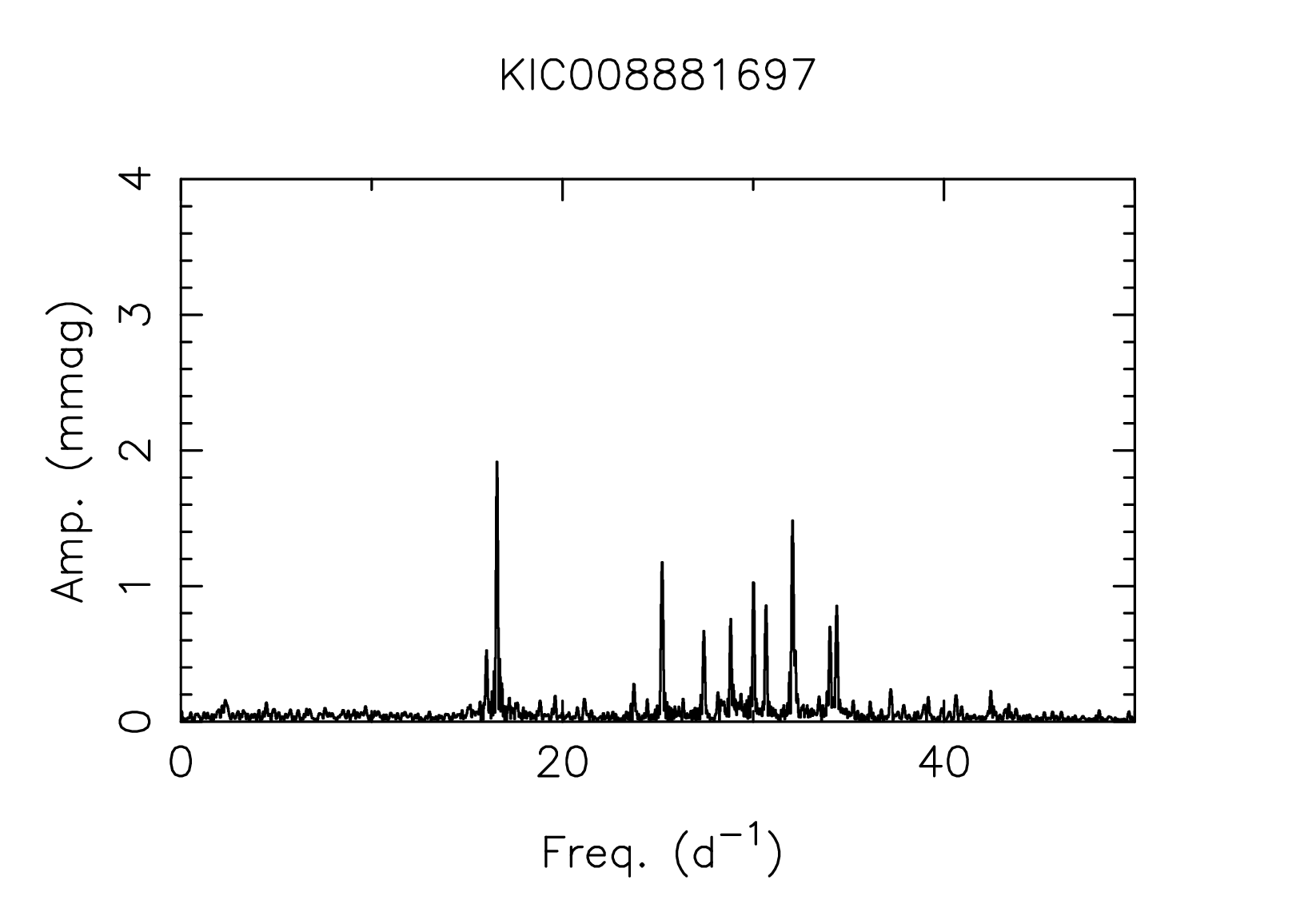}
\includegraphics[height=7cm,width=4.5cm,angle=-90]{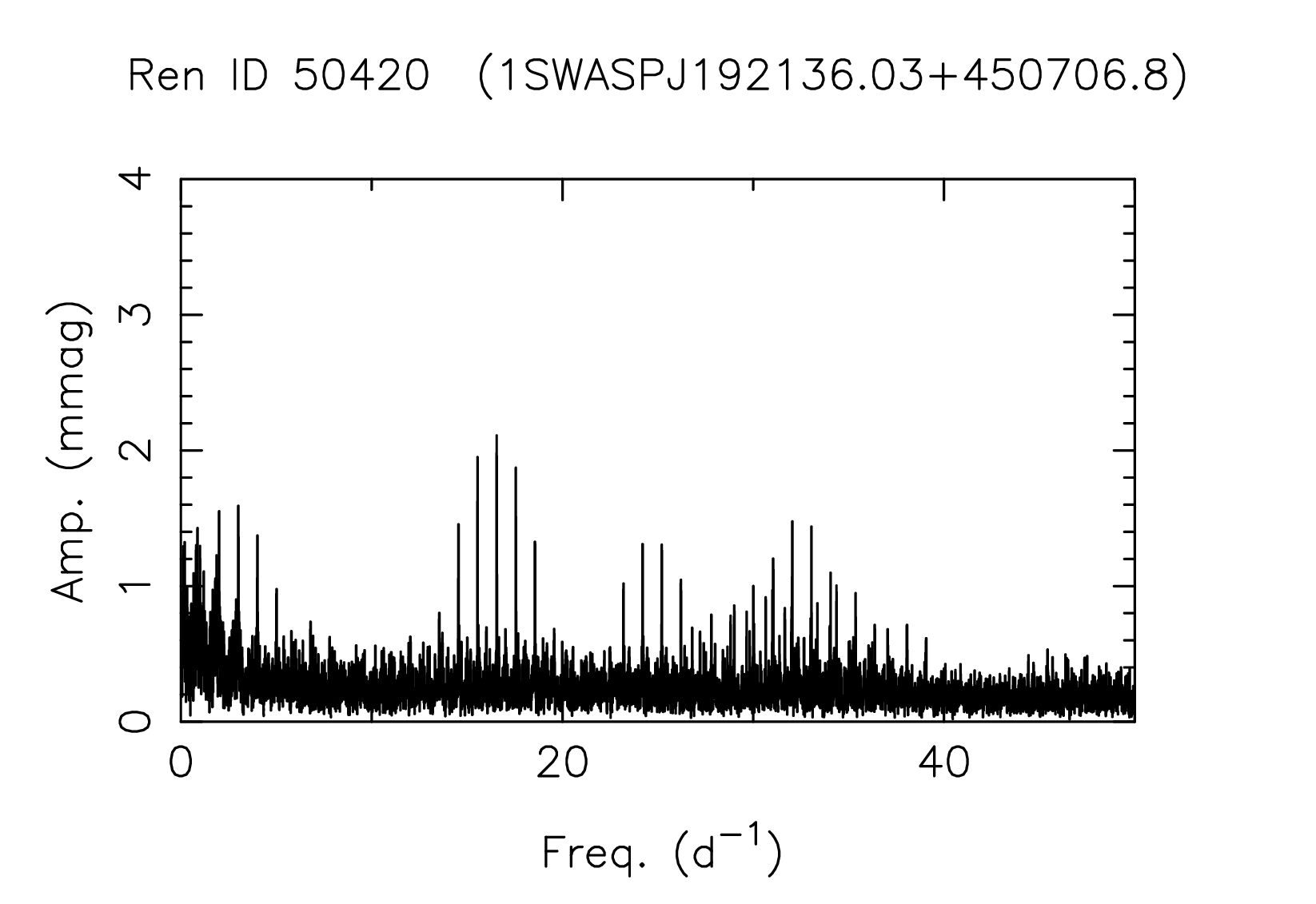}\\
\includegraphics[height=7cm,width=4.5cm,angle=-90]{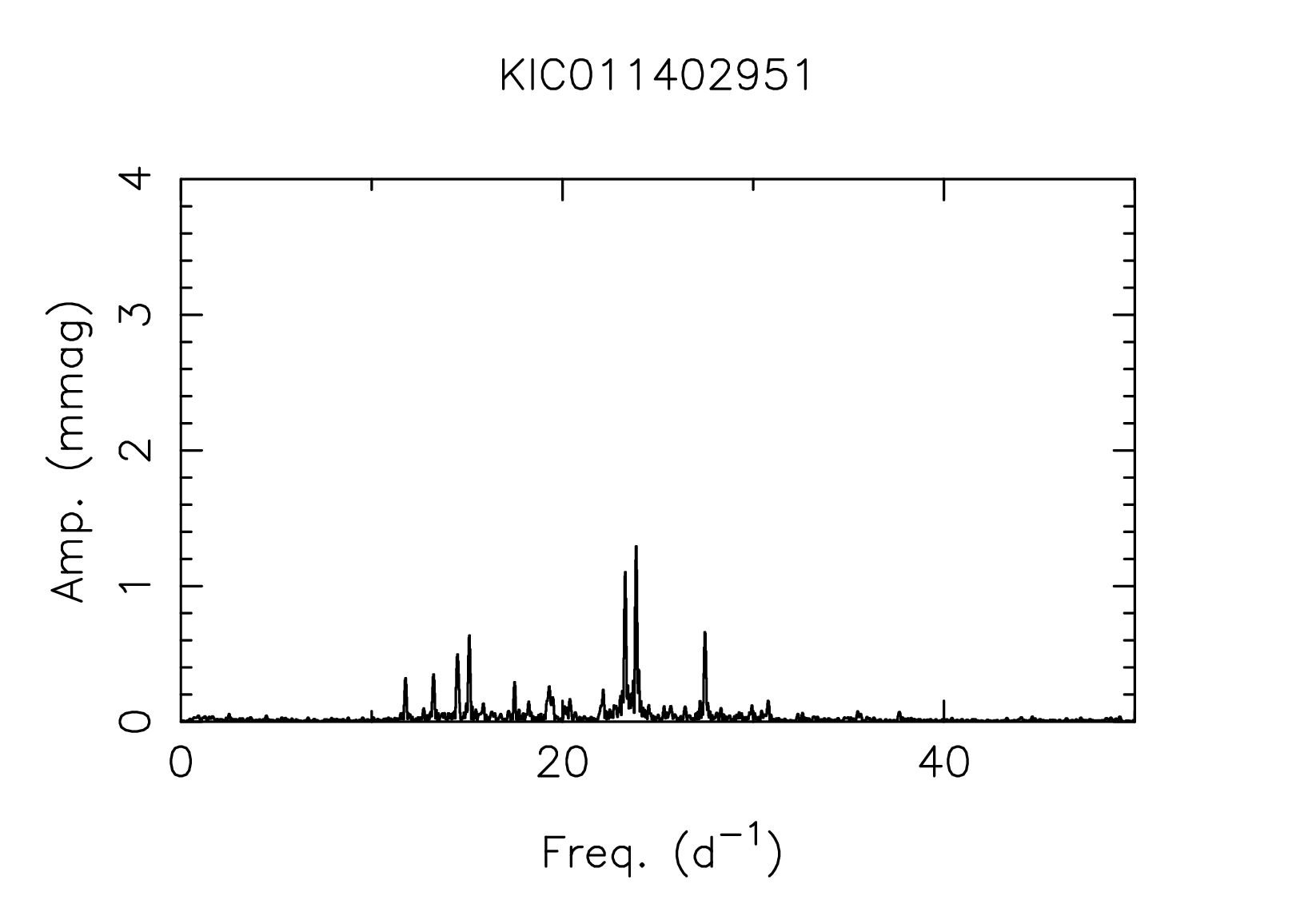}
\includegraphics[height=7cm,width=4.5cm,angle=-90]{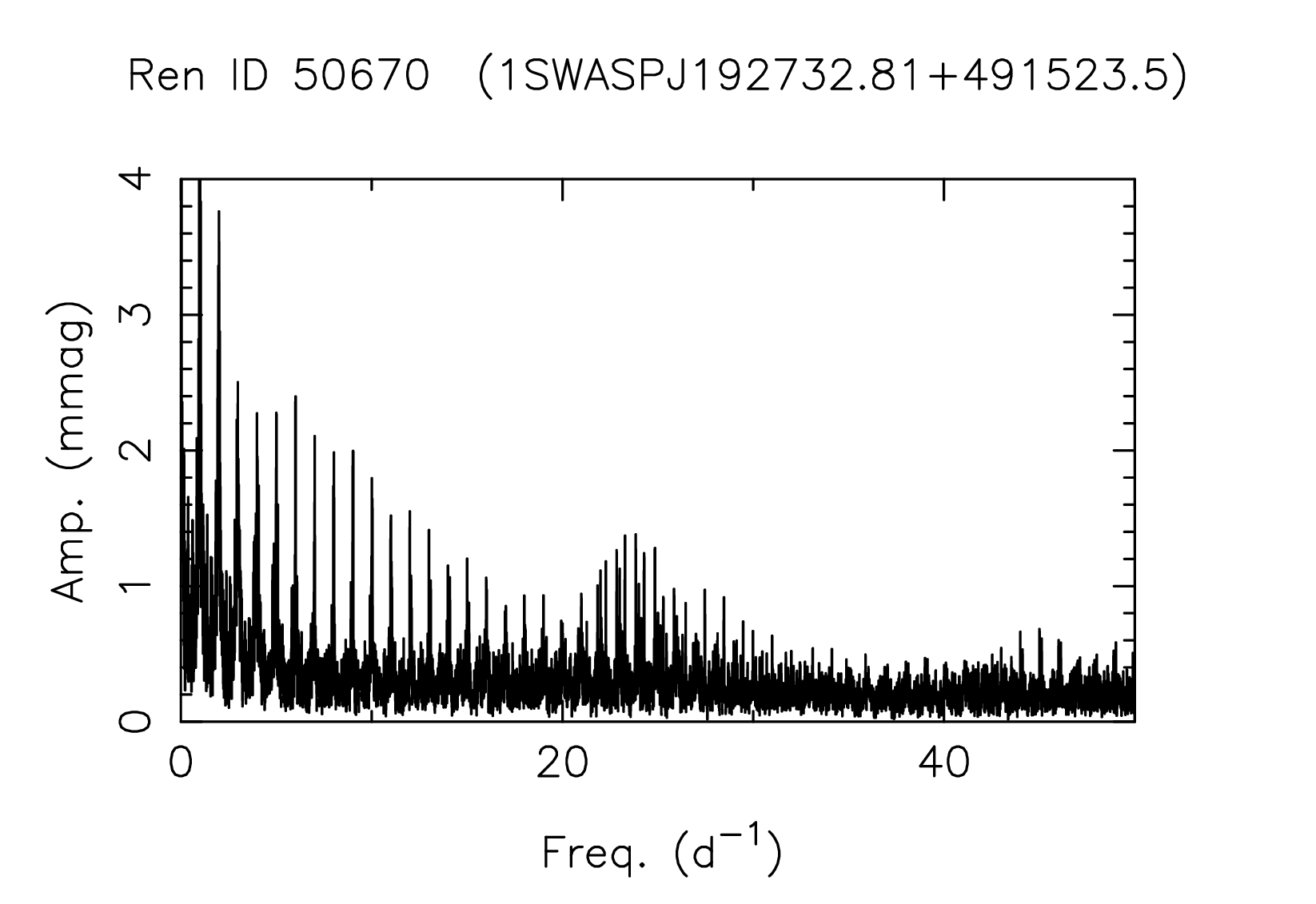}
\caption{Comparison between the {\sc period04} periodograms from {\it Kepler} (left) and
SuperWASP (right) for four Am stars with pulsations detected by SuperWASP (see
Table~\ref{WASP-Kepler-analysis} for details of frequencies identified).}
\label{WASP-Kepler-pdgram}
\end{figure*}

The {\sc period04} analysis (Table~\ref{WASP-Kepler-analysis}) shows good
agreement above the nominal SuperWASP 1\,mmag amplitude limit. There is a
suggestion that the amplitudes found using SuperWASP lightcurves are slightly
higher than those from {\it Kepler}. In addition, the SuperWASP frequency
can differ from the `true' frequency by a small integer number of
1\,d$^{-1}$ aliases. The comparison also shows that it is possible with
SuperWASP data to detect frequencies slightly below the 1\,mmag level
(Figure~\ref{WASP-Kepler-pdgram}). Naturally, the variable data quality of
ground-based photometry means that not all stars with suitable variability will
be detected.

\section{Discussion}

\begin{figure}
\includegraphics[height=\columnwidth,angle=-90]{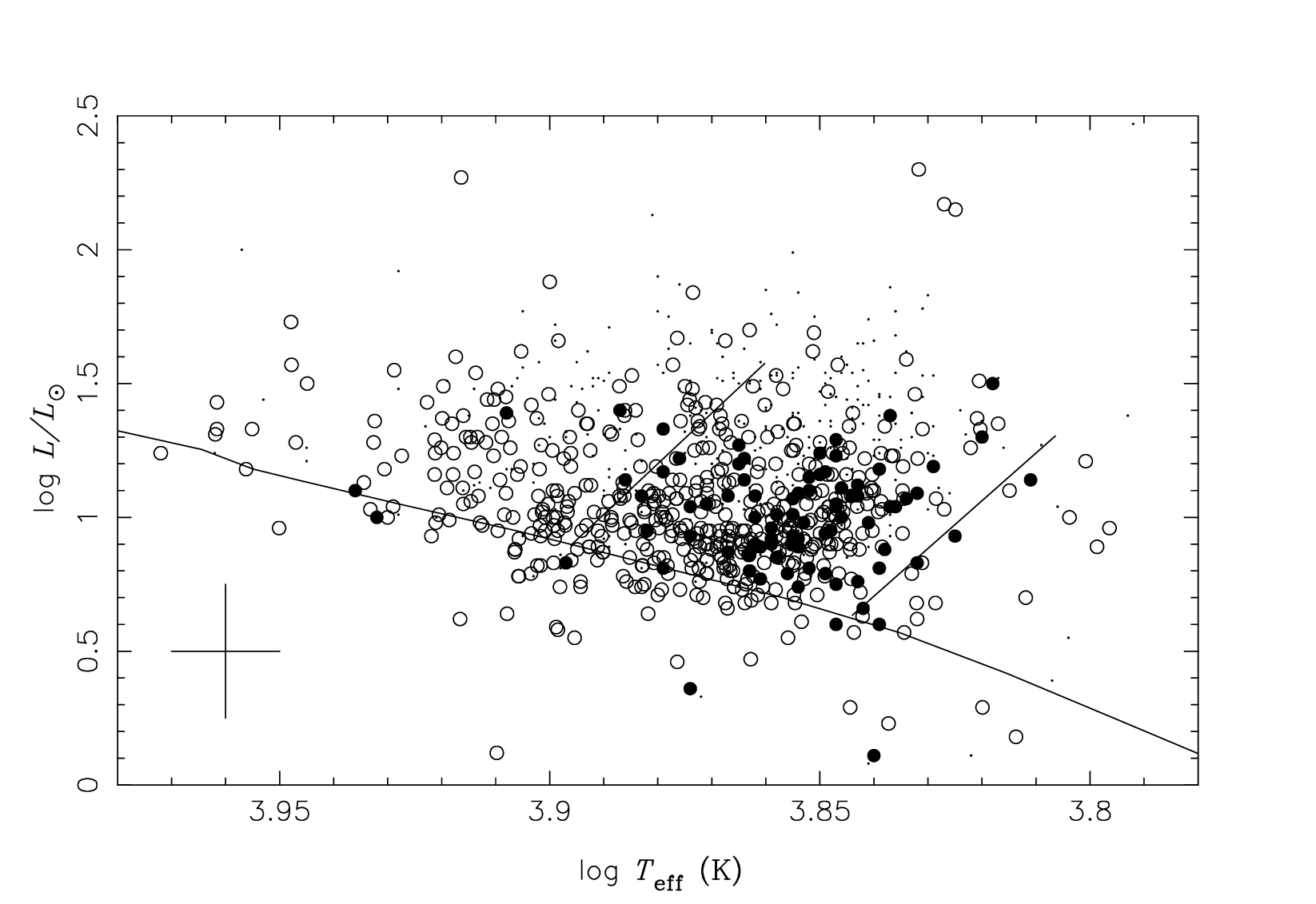}

\caption{HR~diagram showing the location of Am stars. The filled circles are the
Am stars which were found to pulsate, while the open circles are the Am stars
which were not found to pulsate. The solid lines indicate the location of the
ZAMS and the fundamental radial mode red and blue edges of the instability strip
\citep{2005A&A...435..927D}. The large cross indicates the typical uncertainties
in $\log T_\mathrm{eff}$ and $\log L$. The dots are the $\delta$\,Sct stars from
the catalogue of \citet{2000A&AS..144..469R}.}

\label{HR-Diagram}
\end{figure}

The pulsating Am stars (see Fig.\,\ref{HR-Diagram}) are concentrated within the
fundamental radial mode red and blue edges of \cite{2005A&A...435..927D}. This
is in agreement with that found by \cite{2011MNRAS.414..792B} for Am stars
within the {\it Kepler} field. These studies show that pulsating Am stars are
concentrated in the cooler region of the instability strip. Hot Am stars do not
appear to pulsate at the precision of the {\it Kepler} data.

The standard interpretation of the Am phenomenon is that atomic diffusion --
radiative levitation and gravitational settling -- in the outer
stellar envelope gives rise to the observed atmospheric abundance
anomalies. For a typical mid-A star, $T_{\rm eff}\sim8000$~K, there are
two thin convection zones in the outer envelope. The atmosphere itself is a
convection zone a few thousand km thick where ionisation of H drives the
convection. Deeper in the atmosphere, at $T\sim50\,000$~K, the ionisation of
He\,\textsc{ii} also creates a thin convection zone, where the
$\kappa$-mechanism drives $\delta$\,Sct pulsation. It has long been clear that
some Am stars and related types do pulsate, particularly the marginal Am stars
(labelled spectroscopically as Am: stars), the evolved Am stars ($\delta$\,Del
or $\rho$\,Pup stars), and some more extreme cases, such as HD\,188136
(\citealt{1980MNRAS.193...29K}; \citealt{1981ApJ...247..969W}) and HD\,40765
\citep{1995MNRAS.276..199K}.

The pulsation modes that we observe in Am stars are low radial order, low
spherical degree p modes. The surface of the star is an anti-node. With the low
radial order, the vertical wavelength is long compared to the depth of the
envelope above the He\,\textsc{ii} ionisation zone. With the decrease in density
with height in the atmosphere, conservation of kinetic energy density means that
the pulsation amplitude increases with height in the atmosphere, or conversely,
decreases with depth.

In Am stars, the microturbulence velocity is also peculiar, as it is generally
much higher than that of chemically normal stars. This high microturbulence
arises from large velocity fields in the stellar atmosphere
\citep{1998A&A...338.1041L}, which are even supersonic for some Am stars. We do
not really know what causes these large velocity fields to develop exclusively
in Am stars and how chemical peculiarities and velocity fields coexist. The
results shown by \citet{2009A&A...503..973L} suggest that there is a connection
between $T_{\rm eff}$ and the velocity fields, peaking at around $T_{\rm
eff}\sim8000$~K, although we do not know what happens for cooler Am stars.


Atomic diffusion occurs in the radiative zone below the turbulent outer
convective layer, which is far below the observable atmosphere. In this
radiative layer there must be no turbulence at the diffusion velocity, which is
of the order of $10^{-4}$ -- 1~cm\,s$^{-1}$. The photometric amplitudes found in
Am stars are consistent with atmospheric pulsation radial velocity amplitudes of
a few km\,s$^{-1}$. Taking into account the decrease in pulsation amplitude with
depth -- largely because of the increase in density, but also because of the
radial wave function -- the pulsation velocity in the radiative layer where
atomic diffusion is most important in Am stars is still of the order of a
km\,s$^{-1}$. With such pulsations in a layer where atomic diffusion is
operating at sub-cm\,s$^{-1}$ velocities, it must be that the pulsation is
laminar; i.e., producing no turbulence at the sub-cm\,s$^{-1}$ level.

With the results from the {\it Kepler} mission \citep{2011MNRAS.414..792B} and
now our results from SuperWASP we conclude that the loss of helium by
gravitational settling from the He\,\textsc{ii} ionisation zone reduces driving,
but does not suppress it entirely. Thus Am stars can pulsate as $\delta$\,Sct
stars, but typically with relatively low amplitudes compared to normal abundance
$\delta$\,Sct stars. Some Am stars show no pulsation whatsoever at {\it Kepler}
$\mu$mag precision. It has yet to be shown whether this lack of pulsation can
also occur in the more rapidly rotating normal abundance stars in the
$\delta$\,Sct instability strip. Study of this question is in progress with {\it
Kepler} data. As was concluded for the individual cases of HD\,188136 and
HD\,40765, we may now state in general: in Am stars the pulsation must be
laminar, not generating turbulence to mix away the observed effects of atomic
diffusion in the outer atmosphere.

The Fm\,$\delta$\,Del subclass are evolved Am stars above the main-sequence,
many of which have been found to show variability \citep{1976ApJS...32..651K}.
Not unexpectedly, many stars classed as Fm\,$\delta$\,Del are found to be
pulsating in the WASP data, but clearly not all. Of the 227 Am stars that we
found to be pulsating 55 are classed as Fm\,$\delta$\,Del: 24\% of the Am stars
found to pulsate. This compares to a total of 186 Fm\,$\delta$\,Del stars out of
the 1620 Am stars investigated using WASP data, around 11\% of the sample.
Therefore, 30\% of the Fm\,$\delta$\,Del stars have been found to pulsate,
compared to just 12\% of other Am stars. Thus pulsation amplitude either grows
in Am stars as they evolve, or some non-pulsating Am stars begin pulsating as
they move off the main sequence. This is likely to be a consequence of the
driving region moving deeper into the star where the helium abundance is higher
than in the main sequence He\,\textsc{ii} ionisation zone (see
\citealt{2000A&A...360..603T} for theoretical discussion).

\begin{figure}
\includegraphics[height=\columnwidth,angle=-90]{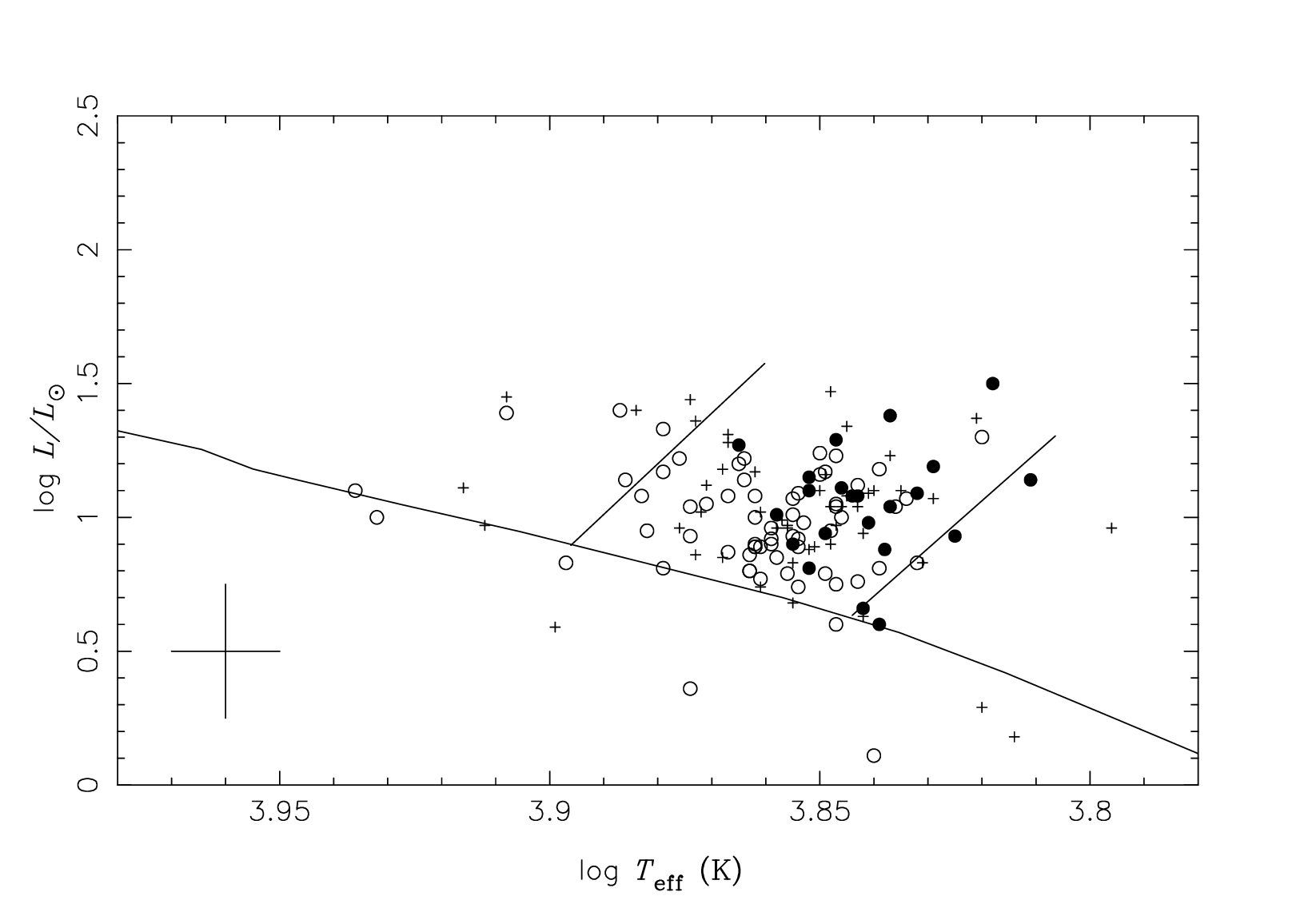}

\caption{Location of the pulsating Am stars in the HR~diagram. The circles are
pulsating Am stars, with the filled circles indicating those with spectral
classification noted as $\delta$\,Del. The crosses are the Fm\,$\delta$\,Del stars
which were not found to pulsate. The solid lines indicate the location of the
ZAMS and the fundamental radial mode red and blue edges of the instability strip
\citep{2005A&A...435..927D}. The large cross indicates the typical uncertainties
in $\log T_\mathrm{eff}$ and $\log L$}

\label{delDel}
\end{figure}

\begin{figure}
\includegraphics[height=\columnwidth,angle=-90]{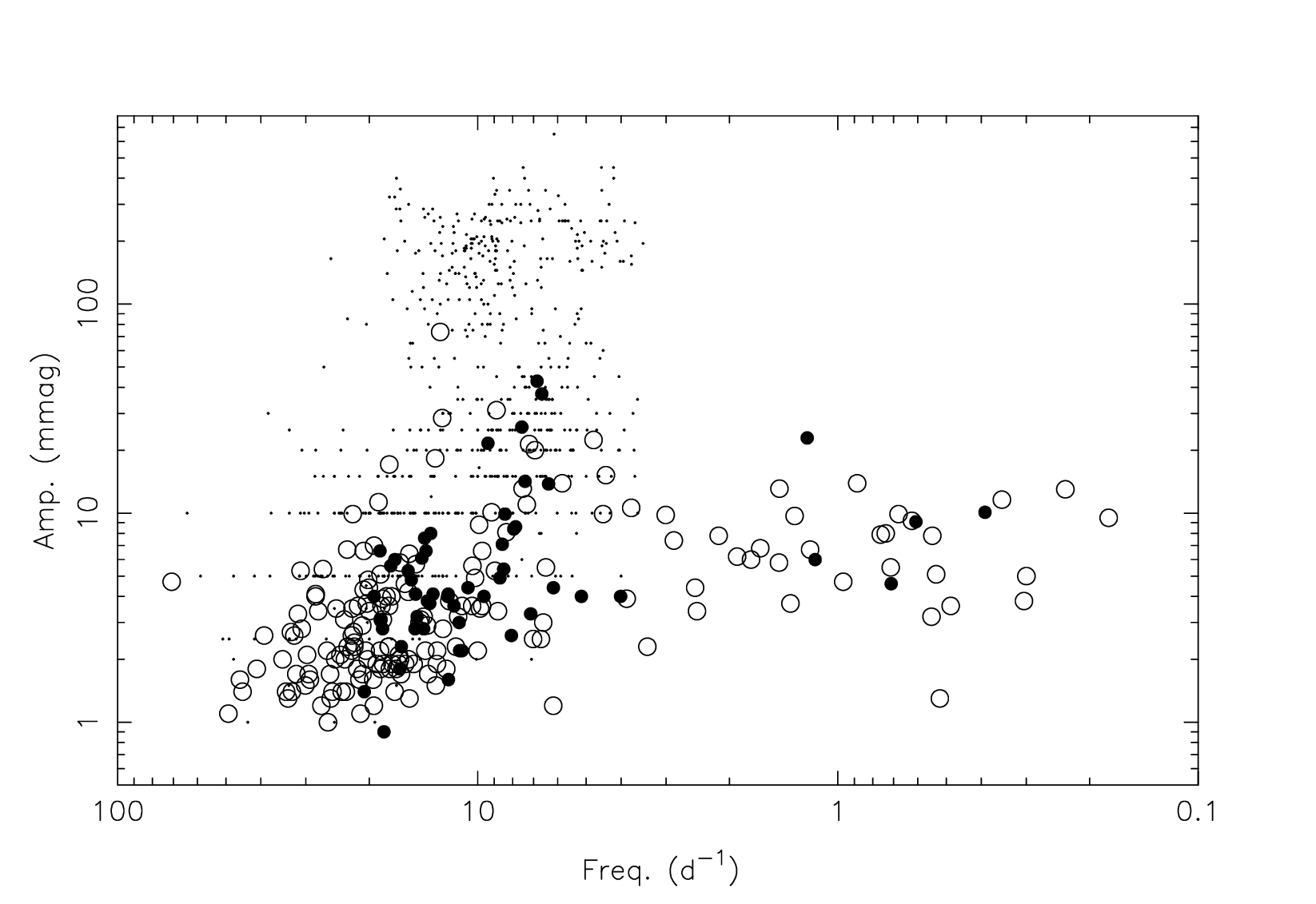}

\caption{Frequency-amplitude diagram for pulsating Am stars shown as circles,
with filled circles indicating those with spectral classification noted as
$\delta$\,Del. Note that in multi-periodic systems only the frequency of the
highest amplitude is shown, as given in Table~\ref{table:Pulsating Am stars}.
The dots are the $\delta$\,Sct stars from the catalogue of
\citet{2000A&AS..144..469R}.}

\label{P-A-delDel}
\end{figure}

The location of the pulsating Fm\,$\delta$\,Del stars in the HR~diagram is shown
in Fig.\,\ref{delDel}. There is a tendency for the pulsating Fm\,$\delta$\,Del
stars to be located toward the cooler (and/or) slightly more evolved parts of
the instability strip, whereas the non-pulsating Fm\,$\delta$\,Del stars are
distributed more uniformly. The frequency--amplitude diagram
(Fig.\,\ref{P-A-delDel}) shows that the Fm\,$\delta$\,Del stars occupy the same
regions as the other Am stars, but with an absence of high-frequency ($\ga
20$\,d$^{-1}$) pulsations; this is not surprising, given that they are cooler
and more evolved than average $\delta$\,Sct stars.

Several factors are thought to play a role in the development of pulsating Am
stars, but stellar rotation is probably one of the most important.
\citet{1991ApJ...370..693C} showed that Am chemical peculiarity develops in
stars that rotate slower than 90\,km\,s$^{-1}$ and that the He\,{\sc ii}
ionisation zone deepens with decreasing rotation. This was later confirmed by
more advanced diffusion model calculations by \citet{2006ApJ...645..634T} and
observationally by \citet{2008A&A...483..891F}, who found a correlation between
Am chemical peculiarities and $v \sin i$ in Am stars belonging to the Praesepe
open cluster. The vast majority of the Am stars already known to pulsate have a
rather large $v \sin i$, between 40 and 90 km\,s$^{-1}$, thus avoiding the
He\,{\sc ii} ionisation zone sinking too deep into the star and therefore
allowing the development of pulsation driven by the $\kappa$-mechanism. On the
other hand, for the very slowly rotating pulsating Am stars, the pulsation could
be laminar. It is therefore likely there are two different mechanisms driving
pulsation in Am stars.

Our results show a wide variety of pulsations, from singly periodic to complex
multiperiodic, and also some examples of what appear to be hybrid
$\gamma$\,Dor/$\delta$\,Sct pulsators. This is similar to the range of behaviour
seen in normal abundance $\delta$\,Sct stars, as can be seen in the study of
{\it Kepler} data by \cite{2010ApJ...713L.192G}. Those authors reclassified
pulsation types with the following scheme:

\begin{description}

\item[{\bf $\delta$\,Sct:}] frequencies above 5\,d$^{-1}$;

\item[{\bf $\delta$\,Sct/$\gamma$\,Dor hybrid:}] most frequencies above 5\,d$^{-
1}$, but some low frequencies present;

\item[{\bf $\gamma$\,Dor:}] frequencies lower than 5\,d$^{-1}$;

\item[{\bf $\gamma$\,Dor/$\delta$\,Sct hybrid:}] most frequencies lower than
5\,d$^{-1}$, but some high frequencies present.

\end{description}

Our results are summarized in Table~\ref{class} and the individual classes for
each star are given in Table~\ref{table:Pulsating Am stars}. The majority of the
pulsators we found are $\delta$\,Sct stars, with the remaining quarter split
between $\gamma$\,Dor stars and mostly $\delta$\,Sct/$\gamma$\,Dor hybrids.
Given that the SuperWASP data are affected by daily aliases and systematics at
low frequencies, the true number of stars with $\gamma$\,Dor pulsations may
indeed be higher. However, given that Am stars are thought to be members of
binary systems and tidal effects slow the stellar rotation rate, it is possible
that some of the low-frequency signatures found in the SuperWASP data are due to
ellipsoidal effects in close binaries. Assuming a rotation limit of $v \sin i
\la 120$\,km\,s$^{-1}$ for an Am star and a radius of 1.5\,$R_{\sun}$, the
shortest period for a binary system containing a tidally-synchronised Am star is
$\sim$0.6\,d. Close binary systems with dissimilar components have two maxima
and minima per orbital period, and this value dominates over the orbital value
in periodograms. Hence, frequencies $\la$3.3\,d$^{-1}$ may have arisen due to
ellipsoidal variations in close binaries. Thus, we caution that some of the
stars presented in Table~\ref{table:Pulsating Am stars} could have erroneously
been classified as having $\gamma$\,Dor pulsations. In addition, it is possible
that long-period pulsations in close binaries could be tidally excited
\citep{2002MNRAS.333..262H}.

It is clear from examination of the {\it Kepler} data set that the $\delta$\,Sct
stars show frequencies ranging from nearly zero d$^{-1}$ up to 100\,d$^{-1}$;
some stars even show the full range, including frequencies between the g\,mode
and p\,mode ranges seen in models. These intermediate frequencies are
unexplained at present. It is clear that the $\delta$\,Sct stars are complex
pulsators that show g\,modes, p\,modes, mixed modes and many nonlinear cross
terms. Whether there are differences between abnormal abundance, slowly rotating
Am stars that are $\delta$\,Sct stars and the more rapidly rotating, normal
abundance $\delta$\,Sct stars is yet to be determined. The objects we present
here from SuperWASP greatly increases the number of pulsating Am stars for
statistical study of this question.

\begin{table}

\caption{The number of pulsating Am stars and percentage in each of the
four pulsation classes as defined by \cite{2010ApJ...713L.192G}.}

\centering
\begin{tabular}{ccc}\hline \hline
Pulsation Class             & Number & Percentage \\\hline
$\delta$\,Sct               &  169   & 75 \\
$\delta$\,Sct/$\gamma$\,Dor &   23   & 10 \\
$\gamma$\,Dor                &   30   & 13 \\
$\gamma$\,Dor/$\delta$\,Sct &    5   &  2 \\
\hline
\end{tabular}
\label{class}
\end{table}

\begin{acknowledgements}

The WASP project is funded and operated by Queen's University Belfast, the
Universities of Keele, St. Andrews and Leicester, the Open University, the Isaac
Newton Group, the Instituto de Astrofisica de Canarias, the South African
Astronomical Observatory and by STFC.
This research has made use of the SIMBAD database,
operated at CDS, Strasbourg, France.
Some of the data presented in this paper were obtained from the Multimission
Archive at the Space Telescope Science Institute (MAST). STScI is operated by the
Association of Universities for Research in Astronomy, Inc., under NASA contract
NAS5-26555. Support for MAST for non-HST data is provided by the NASA Office of
Space Science via grant NNX09AF08G and by other grants and contracts.

\end{acknowledgements}

\longtab{1}{
\small
\begin{longtable}{llllllllllll}
\caption{Pulsating Am stars.}
\label{table:Pulsating Am stars}\\
\hline\hline
Ren ID & Name & Sp. Type & $\log T_\mathrm{eff}$ & $\log L$ & Method & Freq. & Amp. & nFreq. & $\Delta T$ & Class\\
&&&(K)&($L_{\sun}$)&&(d$^{-1}$)&(mmag)&&(d)&\\
\hline
\endfirsthead
\caption{continued.}\\
\hline\hline
Ren ID & Name & Sp. Type & $\log T_\mathrm{eff}$ & $\log L$ & Method & Freq. & Amp. & nFreq. & $\Delta T$ & Class\\
&&&(K)&($L_{\sun}$)&&(d$^{-1}$)&(mmag)&&(d)&\\
\hline
\endhead
\hline
\endfoot
10&HD 154A&A9mF2 &3.862&0.89 &a& 4.7672& 22.4& 2&1135&$\gamma$~Dor\\
110&HD 719&A3mF0 &3.855&1.01 &a& 14.1368& 3.2& 10&473&$\delta$~Sct\\
113&HD 728&A m $\delta$ Del && && 14.9268& 2.8& 2&119&$\delta$~Sct\\
140&HD 923&A6mF2 &3.908&1.39 &a& 18.6684& 2.2& 1&50&$\delta$~Sct\\
210&HD 1097&A4mF4 Sr &3.840&0.11 &a& 15.5491& 2.0& 2&50&$\delta$~Sct\\
355&HD 1651&A6mA9 &3.859&0.92 &a& 15.0600& 1.9& 8&154&$\delta$~Sct\\
500&BD+40 77 A&A2mA7 Sr && && 32.3130& 2.6& 10&146&$\delta$~Sct\\
1233&HD 4630&A3m $\delta$ Del && && 20.6497& 1.4& 1&114&$\delta$~Sct\\
1790&HD 7133&A3 Sr or Am $\delta$ Del ? &3.832&1.09 &a& 14.8755& 4.1& 8&114&$\delta$~Sct\\
1830&CD-22 422&A6m && && 1.9041& 6.2& 4&140&$\gamma$~Dor\\
1920&TYC 5276-1653-1&A6m && && 1.6420& 6.8& 5&140&$\delta$~Sct/$\gamma$~Dor\\
1984&HD 8043&A2mF &3.849&1.17 &a& 10.3662& 3.6& 1&115&$\delta$~Sct\\
2060&HD 8457&A2mF && && 16.5741& 1.9& 7&140&$\delta$~Sct\\
2340&BD-12 290&A2mA7 &3.862&1.08 &b& 34.8219& 2.0& 6&136&$\delta$~Sct\\
2370&HD 9659&A1mA7 &3.883&1.08 &c& 17.6028& 17.1& 8&927&$\delta$~Sct\\
2720&BD+58 304&A7m && && 28.1912& 4.0& 6&115&$\delta$~Sct\\
2920&HD 11490&A5m &3.859&0.96 &a& 6.1679& 1.2& 7&580&$\delta$~Sct\\
3013&TYC 2816-327-1&F1mF4 &3.832&0.83 &a& 0.8838& 13.9& 2&1246&$\gamma$~Dor\\
3340&HD 12961&A5mF3 $\delta$ Del? && && 4.0115& 4.0& 7&563&$\delta$~Sct/$\gamma$~Dor\\
3378&HD 13079&F0m &3.854&0.74 &a& 19.4090& 7.0& 5&480&$\delta$~Sct\\
3550&HD 13776&A0m &3.862&0.90&d& 4.4078& 15.2& 2&563&$\gamma$~Dor\\
3655&HD 14494&A5mA9 &3.850&1.16 &a& 15.5855& 4.2& 10&910&$\delta$~Sct\\
4413&&A3m &3.839&0.81 &a& 21.3019& 1.6& 7&511&$\gamma$~Dor/$\delta$~Sct\\
4793&HD 19108&A3m $\delta$ Del &3.852&0.81 &a& 14.7306& 3.2& 5&121&$\delta$~Sct\\
4885&HD 19762&A5mA8 && && 16.9993& 1.4& 1&145&$\delta$~Sct\\
5055&HD 20308&A5mF0 &3.864&1.14&d& 29.7634& 2.1& 4&521&$\delta$~Sct\\
6044&HD 23543&A5m $\delta$ Del && && 7.1268& 3.3& 3&521&$\delta$~Sct\\
6083&TYC 3729-775-1&A2mF0 && && 24.0686& 2.1& 1&94&$\delta$~Sct\\
6295&TYC 3725-169-1&A3m && && 28.1774& 4.1& 2&1228&$\delta$~Sct\\
6390&HD 25052&A7m &3.853&0.98 &a& 6.5758& 3.0& 7&521&$\delta$~Sct/$\gamma$~Dor\\
6368&HD 24925&A5m $\delta$ Del? && && 18.2088& 0.9& 1&119&$\delta$~Sct\\
6463&HD 25369&A2mA8 && && 31.8867& 1.7& 4&119&$\delta$~Sct\\
6527&HD 25648&A1m &3.865&1.20&d& 1.3183& 9.7& 1&112&$\gamma$~Dor\\
6681&HD 26386&A3mF0 && && 20.7307& 4.3& 7&134&$\delta$~Sct\\
6663&TYC 3726-618-1&A2mF0 && && 22.3602& 2.6& 1&71&$\delta$~Sct\\
8720&HD 34296&A3mF2 &3.861&0.77 &a& 1.7424& 6.0& 9&506&$\delta$~Sct/$\gamma$~Dor\\
8711&HD 242159&A5m && && 8.7898& 3.4& 8&1233&$\delta$~Sct\\
8842&HD 242632A&A5m && && 0.3043& 3.8& 1&924&$\gamma$~Dor\\
8898&HD 34841&A2m &3.886&1.14 &a& 11.4943& 2.3& 4&139&$\delta$~Sct\\
8932&HD 242938&A5m && && 21.6154& 1.8& 3&902&$\delta$~Sct\\
8951&HD 243010&A3m && && 16.4353& 2.0& 2&927&$\delta$~Sct\\
8972&HD 243093&A5m && && 19.9969& 3.4& 6&1232&$\delta$~Sct\\
8974&HD 243112&A2m && && 44.9530& 1.4& 2&139&$\delta$~Sct\\
8988&HD 35236&A2m && && 45.7112& 1.6& 2&139&$\delta$~Sct\\
9084&HD 35467&A2m && && 33.0007& 2.7& 3&1233&$\delta$~Sct\\
9123&HD 243542&A3m && && 25.6983& 1.7& 2&1233&$\delta$~Sct\\
9144&HD 35531&A2m && && 21.4997& 3.6& 5&1232&$\delta$~Sct\\
9262&HD 244020&A7m && && 6.4681& 5.5& 2&1200&$\delta$~Sct\\
9269&&A3m && && 9.7183& 3.6& 2&138&$\delta$~Sct\\
9470&BD-7 1108&A6mF2 &3.854&1.09 &a& 12.7088& 73.6& 4&112&$\delta$~Sct\\
9375&TYC 2411-1663-1&A5m && && 21.9980& 2.4& 5&941&$\delta$~Sct\\
9454&HD 244698&A2m && && 29.4675& 1.7& 2&133&$\delta$~Sct\\
9685&HD 36887&F0m $\delta$ Del? &3.855&0.90 &a& 0.7113& 4.6& 2&152&$\gamma$~Dor\\
9534&HD 244810&A3m && && 18.5025& 3.6& 4&941&$\delta$~Sct\\
9581&HD 36681&A2m &3.858&0.85 &a& 20.8887& 1.7& 1&139&$\delta$~Sct\\
9556&BD+34 1091&A0m && && 1.4571& 5.8& 1&941&$\gamma$~Dor\\
9653&HD 245063&A3m && && 11.0583& 3.6& 1&941&$\delta$~Sct\\
9812&HD 245303&A3m && && 17.6418& 3.6& 5&138&$\delta$~Sct\\
9868&&A3m && && 0.6786& 9.9& 3&152&$\gamma$~Dor\\
10206&TYC 1869-592-1&A5m && && 23.4739& 3.1& 1&133&$\delta$~Sct\\
10259&HD 246984&A5m && && 20.7162& 6.6& 3&935&$\delta$~Sct\\
10383&HD 247634&A7m && && 2.8538& 7.4& 3&928&$\gamma$~Dor\\
10423&HD 247837&A1m && && 0.2336& 13.0& 2&923&$\gamma$~Dor\\
10610&HD 39641&A5mF1 &3.863&0.86 &a& 20.4628& 2.2& 8&177&$\delta$~Sct/$\gamma$~Dor\\
10448&HD 248069&A0m && && 3.7473& 10.6& 2&923&$\gamma$~Dor/$\delta$~Sct\\
10474&HD 248174&A2m && && 18.3207& 1.9& 1&133&$\delta$~Sct\\
10482&HD 248244&A1m && && 49.2363& 1.1& 1&138&$\delta$~Sct\\
10565&HD 248637&A2m && && 12.4952& 2.8& 5&133&$\delta$~Sct\\
10554&HD 248577&A0m && && 11.3345& 3.2& 1&1194&$\delta$~Sct\\
10586&TYC 1867-814-1&A5m && && 12.5497& 28.5& 5&923&$\delta$~Sct/$\gamma$~Dor\\
10602&HD 248874&A0m && && 0.1771& 9.5& 1&138&$\gamma$~Dor\\
10641&HD 249278&A5m && && 29.2307& 1.6& 1&138&$\delta$~Sct\\
10855&TYC 1876-325-1&A0m && && 9.9108& 8.8& 1&138&$\delta$~Sct\\
11007&HD 251038&A0m && && 14.3993& 3.1& 3&1915&$\delta$~Sct\\
11022&HD 251095&A2m && && 7.3109& 11.0& 6&1218&$\delta$~Sct\\
11025&HD 251143&A3m && && 17.2563& 1.9& 3&133&$\delta$~Sct\\
11032&HD 251227&A3m && && 3.8590& 3.9& 4&133&$\delta$~Sct/$\gamma$~Dor\\
11033&HD 251226&A5m && && 22.1084& 2.7& 2&133&$\delta$~Sct\\
11135&HD 251963&A7m && && 10.1655& 4.9& 4&1218&$\delta$~Sct\\
11184&HD 252154&A5m && && 1.4525& 13.1& 5&919&$\delta$~Sct/$\gamma$~Dor\\
11182&TYC 1889-117-1&A3m && && 6.6767& 2.5& 1&133&$\delta$~Sct\\
11274&HD 252679&A3mF2 && && 30.1049& 1.5& 2&138&$\delta$~Sct\\
11265&HD 42155&A2m && && 22.0073& 2.3& 4&133&$\delta$~Sct\\
11473&TYC 1314-887-1&A3mF2 && && 1.1943& 6.7& 1&138&$\gamma$~Dor\\
11820&HD 44596&A6mF2 &3.847&1.04 &a& 11.9914& 3.8& 15&177&$\delta$~Sct\\
12250&HD 45863&A2mA8 &3.879&0.81 &c& 0.5209& 1.3& 1&1040&$\gamma$~Dor\\
12820&HD 47743&A3mF0 &3.863&0.80 &a& 18.3705& 3.9& 3&66&$\gamma$~Dor/$\delta$~Sct\\
12940&HD 48223&A4m $\delta$ Del? && && 18.5331& 3.1& 13&795&$\delta$~Sct/$\gamma$~Dor\\
13550&CP-60 704&F0m && && 14.5060& 3.0& 9&177&$\delta$~Sct\\
14140&HD 51319&A2mA9 && && 15.9002& 1.9& 5&177&$\delta$~Sct\\
15423&HD 56484&A2mF0 &3.876&1.22 &a& 14.8533& 5.7& 3&132&$\delta$~Sct\\
16920&HD 61659&A5mF0 &3.874&0.93 &a& 0.7136& 5.5& 1&142&$\gamma$~Dor\\
18657&HD 67518&A3m $\delta$ Del &3.841&0.98 &a& 5.1544& 4.0& 2&117&$\delta$~Sct/$\gamma$~Dor\\
18730&HD 67911&F0mF4 &3.834&1.07 &a& 8.8671& 31.1& 3&98&$\delta$~Sct\\
20143&HD 72658&F m $\delta$ Del? && && 8.6775& 4.9& 1&96&$\delta$~Sct\\
20333&HD 73144&F m $\delta$ Del? && && 8.0629& 2.6& 1&97&$\delta$~Sct\\
20485&HD 73675&A4m && && 33.6236& 1.3& 4&96&$\delta$~Sct\\
20855&HD 74626&A5m $\delta$ Del && && 7.8486& 8.6& 10&119&$\delta$~Sct\\
20908&HD 74784&A3mF2 $\delta$ Del? &3.837&1.38 &a& 7.5427& 25.8& 12&97&$\delta$~Sct\\
21730&HD 77105&A3m Sr $\delta$ Del &3.829&1.19&d& 11.2579& 3.0& 6&119&$\delta$~Sct\\
21920&HD 77532&A5mF0 && && 19.5074& 1.6& 4&119&$\delta$~Sct\\
22180&HD 78325&A2mA8 && && 7.0172& 2.5& 5&717&$\delta$~Sct/$\gamma$~Dor\\
22404&HD 79034A&F0m $\delta$ Del && && 14.3002& 6.1& 2&96&$\delta$~Sct\\
22450&HD 79111&A2mF2 && && 15.9320& 4.6& 1&745&$\delta$~Sct\\
22685&HD 79787&A3mF2 && && 20.9176& 2.9& 4&749&$\delta$~Sct\\
23195&HD 81729&A5m &3.850&1.24&d& 9.7198& 6.6& 4&131&$\delta$~Sct\\
23410&HD 82396&F0m $\delta$ Del &3.847&1.29 &a& 9.3672& 21.6& 6&110&$\delta$~Sct\\
23672&HD 83049&F5m $\delta$ Del && && 7.3912& 14.2& 3&511&$\delta$~Sct\\
24920&HD 87118&A4mF2 && && 0.5334& 5.1& 1&521&$\gamma$~Dor\\
24990&HD 87360&F0m $\delta$ Del &3.865&1.27 &a& 13.5113& 8.0& 2&457&$\delta$~Sct\\
25160&HD 87869&A2m $\delta$ Del && && 13.8008& 3.8& 7&113&$\delta$~Sct\\
25730&BD+42 2113&A mF && && 8.3255& 8.1& 2&134&$\delta$~Sct\\
26360&HD 91616&A3mF3 && && 0.4861& 3.6& 1&717&$\gamma$~Dor\\
26860&HD 93038&A3m && && 17.9074& 4.0& 7&717&$\delta$~Sct/$\gamma$~Dor\\
26880&HD 93137&F5m $\delta$ Del && && 6.6384& 37.2& 7&1090&$\delta$~Sct\\
27270&HD 94479&A4mF0 &3.862&1.00 &a& 20.1477& 4.8& 3&123&$\delta$~Sct\\
27405&HD 95192&A1mF0 && && 19.0772& 1.9& 2&478&$\delta$~Sct\\
27526&HD 95562&A2mA9 &3.871&1.05&d& 41.0464& 1.8& 1&144&$\delta$~Sct\\
27610&HD 95856&F5m $\delta$ Del &3.843&1.08 &a& 17.4329& 5.6& 6&510&$\delta$~Sct\\
28290&HD 98009&A3mF0 && && 34.0300& 1.4& 5&510&$\delta$~Sct\\
28340&HD 98299&A3mA7 && && 13.0735& 1.5& 2&144&$\delta$~Sct\\
28510&HD 98946&A5m &3.887&1.40 &a& 18.6186& 5.1& 3&1097&$\delta$~Sct\\
28610&HD 99302&A3mF1 &3.882&0.95 &b& 39.1743& 2.6& 2&134&$\delta$~Sct\\
28690&HD 99729&F5m $\delta$ Del &3.825&0.93&d& 7.9438& 8.4& 9&510&$\delta$~Sct\\
28850&HD 100376&F0m $\delta$ Del? && && 1.2169& 22.9& 3&510&$\gamma$~Dor\\
29280&BD+41 2224&A0m && && 0.3505& 11.6& 2&1108&$\gamma$~Dor\\
29310&HD 101717&A5m $\delta$ Del && && 18.3663& 2.8& 2&142&$\delta$~Sct\\
29590&HD 102594&F2m $\delta$ Del && && 16.3003& 2.3& 9&414&$\delta$~Sct\\
29800&HD 103318&A4m && && 16.3386& 1.7& 2&132&$\delta$~Sct\\
30390&HD 104957&A3mF1 &3.867&0.87 &a& 18.8594& 11.3& 6&409&$\gamma$~Dor/$\delta$~Sct\\
30453&BD+18 2569&A m &3.847&0.75 &a& 16.8163& 1.8& 5&1058&$\delta$~Sct\\
30926&HD 106832A&A1mA9 Hg && && 23.3836& 2.0& 1&496&$\delta$~Sct\\
30970&TYC 2530-1366-1&A2m && && 10.0009& 2.2& 1&1108&$\delta$~Sct\\
31500&HD 108452&A0m && && 70.7600& 4.7& 7&117&$\delta$~Sct/$\gamma$~Dor\\
31560&HD 108668&A3mF2 && && 23.2410& 1.4& 1&153&$\delta$~Sct\\
31600&BD+37 2284&A7mF3 &3.879&1.17 &a& 22.2298& 9.9& 8&1123&$\delta$~Sct/$\gamma$~Dor\\
31680&&A5m &3.874&1.04 &b& 2.1429& 7.8& 4&1123&$\delta$~Sct/$\gamma$~Dor\\
31710&HD 109306&F2m &3.836&1.04 &a& 3.0068& 9.8& 3&136&$\delta$~Sct/$\gamma$~Dor\\
31800&TYC 2533-2112-1&A2m && && 24.6919& 3.5& 1&136&$\delta$~Sct\\
31913&HD 109957&A3m $\delta$ Del &3.839&0.60&d& 13.3080& 4.1& 2&509&$\delta$~Sct\\
31950&HD 110056&A3mF0 &3.847&0.60&d& 16.4269& 5.8& 10&755&$\delta$~Sct\\
32180&BD+38 2361&A6mF2 &3.864&1.22 &a& 20.0646& 4.4& 3&1123&$\delta$~Sct\\
32340&BD+21 2457&A3m &3.847&1.05 &a& 13.1200& 18.3& 8&136&$\delta$~Sct\\
32624&HD 112340&A2m &3.936&1.10 &a& 31.0376& 5.3& 3&779&$\delta$~Sct\\
32870&HD 113221&A3mF0 &3.854&0.89 &c& 25.5893& 1.3& 2&713&$\delta$~Sct\\
32885&HD 113385&F0m $\delta$ Del &3.842&0.66 &a& 0.3905& 10.1& 5&483&$\gamma$~Dor\\
33220&HD 114839&A3m &3.863&0.80 &a& 0.9665& 4.7& 9&1080&$\delta$~Sct/$\gamma$~Dor\\
33490&HD 115800A&F5m $\delta$ Del && && 6.1654& 4.4& 1&167&$\delta$~Sct\\
33555&HD 116276&F0m $\delta$ Del &3.846&1.11 &a& 10.6404& 4.4& 6&1141&$\delta$~Sct\\
33636&HD 116635&A2m &3.932&1.00 &a& 0.2997& 5.0& 2&136&$\gamma$~Dor\\
33940&HD 117682&F5m $\delta$ Del && && 13.9189& 6.6& 11&713&$\delta$~Sct\\
34137&BD+35 2465&A m &3.897&0.83 &a& 0.7607& 7.9& 1&1138&$\gamma$~Dor\\
34076&HD 118209&A3mF3 && && 17.6727& 2.3& 7&396&$\delta$~Sct\\
34620&HD 120054&A2mA8 && && 1.3539& 3.7& 2&713&$\gamma$~Dor/$\delta$~Sct\\
34930&HD 121352&A4mF3 && && 15.4849& 6.4& 2&483&$\delta$~Sct\\
34920&HD 121290&A2mA9 && && 0.7363& 8.0& 1&755&$\gamma$~Dor\\
34996&HD 121698&A4mF2 && && 0.5498& 3.2& 1&143&$\gamma$~Dor\\
35074&HD 122370&F0m $\delta$ Del &3.849&0.94 &a& 15.5955& 5.3& 2&489&$\delta$~Sct\\
35450&HD 123937&F2m $\delta$ Del? && && 6.8438& 42.8& 11&714&$\delta$~Sct\\
35490&HD 124028&F2m $\delta$ Del? && && 11.0526& 2.2& 1&752&$\delta$~Sct\\
35650&HD 124467&A2mF2 && && 26.8987& 5.4& 5&752&$\delta$~Sct\\
35710&HD 124891&A3m $\delta$ Del && && 1.1567& 6.0& 2&714&$\gamma$~Dor\\
35776&HD 125296&A4mF0 && && 20.2108& 2.0& 2&143&$\delta$~Sct/$\gamma$~Dor\\
36080&HD 126685&A5mF2 && && 17.3456& 4.0& 3&755&$\delta$~Sct\\
36330&HD 127832&F5m $\delta$ Del && && 8.4714& 5.4& 5&752&$\delta$~Sct\\
36940&HD 129570&A2mF0 && && 26.1936& 2.2& 3&153&$\delta$~Sct\\
37513&HD 132092&A5mF0 &3.862&0.89 &a& 18.3348& 3.1& 4&464&$\delta$~Sct\\
37494&HD 132054&A3mA9 && && 17.5428& 1.8& 1&890&$\delta$~Sct\\
37884&HD 133489&A2m $\delta$ Del && && 9.6097& 4.0& 8&1178&$\delta$~Sct\\
38400&HD 135306&F5m $\delta$ Del && && 12.1156& 4.0& 7&753&$\delta$~Sct\\
40280&HD 141976&A6m && && 22.3807& 2.2& 4&167&$\delta$~Sct\\
40613&HD 143439&F0m $\delta$ Del &3.852&1.10 &a& 14.1303& 2.8& 5&1192&$\delta$~Sct\\
40675&HD 143517&A3 Sr or $\delta$ Del && && 14.0384& 7.6& 18&1192&$\delta$~Sct/$\gamma$~Dor\\
40805&HD 144033&A3mF4 &3.779&2.72 &b& 15.4669& 1.3& 2&391&$\delta$~Sct\\
41030&HD 144768&A5m $\delta$ Del && && 16.9885& 6.0& 2&657&$\delta$~Sct\\
41315&HD 146053&A6mF3 && && 9.8533& 3.5& 14&1192&$\delta$~Sct/$\gamma$~Dor\\
41640&HD 147400&A2mF3 && && 12.2295& 1.8& 6&755&$\delta$~Sct\\
43588&HD 154225&A5m &3.820&1.30 &b& 6.9346& 20.0& 4&1047&$\delta$~Sct\\
43590&HD 154226&A2m &3.879&1.33&d& 0.6233& 9.2& 1&127&$\gamma$~Dor\\
45870&BD+46 2371&A3m &3.861&0.89 &a& 32.7894& 1.4& 5&126&$\delta$~Sct\\
46050&BD+45 2607&A3m && && 13.7280& 1.7& 4&126&$\delta$~Sct\\
49650&HD 178327&A6mF5 &3.859&0.90 &b& 31.5577& 3.3& 3&456&$\delta$~Sct\\
50230&HD 181206&A5m && && 27.1776& 1.2& 3&128&$\delta$~Sct\\
50420&BD+44 3115&A5m && && 16.5565& 2.1& 7&128&$\delta$~Sct\\
50670&HD 183489&A9mF5 &3.855&1.07 &b& 23.8464& 1.4& 3&1562&$\delta$~Sct\\
50520&HD 182684&A7m $\delta$ Del? && && 12.1003& 4.0& 12&519&$\delta$~Sct\\
51760&HD 187698&A3mF0 && && 0.5468& 7.8& 1&755&$\gamma$~Dor\\
52850&HD 190242&A1mF2 &3.854&0.92 &c& 26.0473& 1.0& 2&755&$\delta$~Sct\\
54150&HD 193981&A6m && && 24.8629& 2.0& 8&162&$\delta$~Sct\\
54515&HD 195638&F2m $\delta$ Del &3.811&1.14&d& 8.4041& 9.9& 1&124&$\delta$~Sct\\
54656&HD 196100&A3mA9 && && 27.7231& 3.4& 2&890&$\delta$~Sct\\
54736&HD 196414&A3m $\delta$ Del &3.838&0.88&d& 11.2288& 2.2& 3&890&$\delta$~Sct\\
54970&HD 197105&A5m &3.856&0.79 &a& 7.5023& 13.1& 2&832&$\delta$~Sct\\
55087&HD 235334&A mF && && 8.9589& 5.3& 3&135&$\delta$~Sct\\
55094&HD 197778&A2m $\delta$ Del? && && 15.2904& 4.8& 4&105&$\delta$~Sct\\
55710&HD 200057&A4mF $\delta$ Del && && 18.6734& 6.6& 10&907&$\delta$~Sct\\
56159&BD+37 4187&A5m && && 12.9635& 2.2& 6&125&$\delta$~Sct\\
56110&HD 201150A&A3mF0 &3.843&0.76 &c& 17.6571& 2.3& 2&907&$\delta$~Sct\\
56275&BD+34 4321&A7 Si Sr or A5m && && 5.8258& 13.9& 7&125&$\delta$~Sct\\
56280&HD 201816&A3mF0 &3.874&0.36&d& 22.2879& 3.5& 3&179&$\delta$~Sct\\
56770&HD 203880&A5mA9 &3.848&0.95 &c& 12.9823& 1.9& 4&179&$\delta$~Sct\\
56980&HD 204620&A9m && && 10.3385& 5.6& 3&537&$\delta$~Sct\\
57020&HD 204806&A4mF1 && && 2.4913& 4.4& 4&536&$\delta$~Sct/$\gamma$~Dor\\
57104&HD 204972&A2mF2 &3.849&0.79&d& 13.8158& 2.9& 4&1219&$\delta$~Sct\\
57300&HD 205651&A4mF3 $\delta$ Del? && && 11.6332& 3.6& 13&536&$\delta$~Sct\\
57323&HD 205813&F0m $\delta$ Del &3.858&1.01 &a& 16.4598& 1.8& 4&141&$\delta$~Sct/$\gamma$~Dor\\
57696&TYC 3975-745-1&A5m && && 13.9770& 2.2& 5&140&$\delta$~Sct\\
57760&HD 207658&A5m $\delta$ Del? && && 19.3944& 4.0& 8&537&$\delta$~Sct\\
57764&HD 207723&A1mF3 && && 30.8011& 2.8& 4&439&$\delta$~Sct\\
58270&HD 209430A&A2mF2 &3.839&1.18 &c& 22.9653& 2.3& 5&556&$\delta$~Sct\\
58440&HD 209930&A3m $\delta$ Del &3.844&1.08 &a& 13.5892& 3.7& 1&179&$\delta$~Sct\\
58850&HD 212108&A2m &3.867&1.08&d& 19.4303& 1.2& 4&112&$\delta$~Sct\\
58870&HD 212164&A1mA9 && && 2.4622& 3.4& 5&556&$\delta$~Sct/$\gamma$~Dor\\
59072&TYC 3611-1607-1&A3m && && 4.4864& 9.9& 2&120&$\gamma$~Dor\\
59020&HD 212765&A2mF2 && && 23.0318& 6.7& 8&556&$\delta$~Sct\\
59090&HD 213204&F1m &3.843&1.12 &a& 9.1521& 10.1& 4&549&$\delta$~Sct\\
59500&HD 215396&A2mF3 &3.846&1.00 &c& 18.5845& 1.8& 5&555&$\delta$~Sct\\
59560&HD 215611&A8mF3 $\delta$ Del &3.852&1.15 &a& 6.3584& 13.8& 4&1218&$\delta$~Sct\\
60696&HD 221446&F2m $\delta$ Del && && 12.0775& 4.1& 3&141&$\delta$~Sct\\
60690&HD 221431&A5m $\delta$ Del &3.837&1.04 &b& 0.6072& 9.1& 2&563&$\gamma$~Dor\\
60740&HD 221576&A2m && && 25.3063& 1.4& 1&506&$\delta$~Sct\\
61105&HD 222828&F2m $\delta$ Del &3.818&1.50 &b& 8.5454& 7.1& 2&546&$\delta$~Sct\\
61320&HD 223676&A2mA8 &3.855&0.93 &c& 21.1612& 1.1& 1&563&$\delta$~Sct\\
61350&BD+44 4512&A2mF5 && && 3.3802& 2.3& 1&140&$\gamma$~Dor\\
61356&HD 223944&A7mF4 && && 7.1976& 21.4& 6&154&$\delta$~Sct\\
61580&HD 224657&F0mF5 &3.847&1.23 &b& 20.3278& 3.7& 2&140&$\delta$~Sct\\
61756&HD 225184&A2m $\delta$ Del && && 12.0588& 1.6& 7&520&$\delta$~Sct\\
\hline
\end{longtable}
\tablefoot{
The first column gives the identification number (Ren ID) from the \cite{2009A&A...498..961R} catalogue.
In column 6, the method of stellar parameter determination is given:
a) $uvby\beta$ photometry, b) $uvby$ photometry
c) Geneva photometry, d) spectral energy distribution and parallax.
Freq. is the frequency of the highest amplitude (Amp.) and nFreq is the number of identified frequencies.
$\Delta T$ is the time baseline of the SuperWASP photometry.
Class is the pulsation class as defined by \cite{2010ApJ...713L.192G}.
}
}


\begin{thebibliography}{}

\bibitem[Abt(2009)]{2009AJ....138...28A}
Abt, H.~A. 2009, \aj, 138, 28

\bibitem[Aerts et al.(2010)]{2010aste.book.....A}
Aerts, C., Christensen-Dalsgaard, J., \& Kurtz, D.~W. 2010,
Asteroseismology (Springer, Heidelberg)

\bibitem[Balona et al.(2011)]{2011MNRAS.414..792B}
Balona, L., Ripepi, V., Catanzaro, G., et al. 2011, \mnras, 414, 792

\bibitem[Breger et al.(1993)]{1993A&A...271..482B}
Breger, M., Stich, J., Garrido, R., et al. 1993, \aap, 271, 482

\bibitem[Butters et al.(2010)]{2010A&A...520L..10B}
Butters, O.~W., West, R.~G., Anderson, D.~R., et al. 2010, \aap, 520, L10

\bibitem[Catanzaro et al.(2011)]{2011MNRAS.411.1167C}
Catanzaro, G., Ripepi, V., Bernabei, S., et al. 2011, \mnras, 411, 1167

\bibitem[Charbonneau \& Michaud(1991)]{1991ApJ...370..693C}
Charbonneau, P., \& Michaud, G. 1991, \apj, 370, 693

\bibitem[Dupret et al.(2005)]{2005A&A...435..927D}
Dupret, M.-A., Grigahc{\`e}ne, A., Garrido, R., Gabriel, M., \& Scuflaire, R.
2005, \aap, 435, 927

\bibitem[Fossati et al.(2008)]{2008A&A...483..891F} 
Fossati, L., Bagnulo, S., Landstreet, J., et al. 2008, \aap, 483, 891

\bibitem[Gilliland et al.(2010)]{2010PASP..122..131G}
Gilliland, R.~L., Brown, T.~M., Christensen-Dalsgaard, J., et al. 2010, \pasp,
122, 131

\bibitem[Grigahc{\`e}ne et al.(2010)]{2010ApJ...713L.192G}
Grigahc{\`e}ne, A., Antoci, V., Balona, L., et al. 2010, \apjl, 713, L192

\bibitem[Handler et al.(2002)]{2002MNRAS.333..262H}
Handler, G., Balona, L.~A., Shobbrook, R.~R., et al. 2002, \mnras, 333, 262

\bibitem[Hauck \& Mermilliod(1998)]{1998A&AS..129..431H}
Hauck, B., \& Mermilliod, M. 1998, \aaps, 129, 431

\bibitem[Henry \& Fekel(2005)]{2005AJ....129.2026H}
Henry, G.~W., \& Fekel, F.~C. 2005, \aj, 129, 2026

\bibitem[Horne \& Baliunas(1986)]{1986ApJ...302..757H}
Horne, J.~H., \& Baliunas, S.~L. 1986, \apj, 302, 757

\bibitem[Joshi et al.(2006)]{2006A&A...455..303J}
Joshi, S., Mary, D.~L., Martinez, P., et al. 2006, \aap, 455, 303

\bibitem[Joshi et al.(2009)]{2009A&A...507.1763J}
Joshi, S., Mary, D.~L., Chakradhari, N.~K., Tiwari, S.~K., \& Billaud, C.
2009, \aap, 507, 1763

\bibitem[K\"unzli et al.(1997)]{1997A&AS..122...51K}
K\"unzli, M., North, P., Kurucz, R.~L., \& Nicolet, B. 1997, \aaps, 122, 51

\bibitem[Kurtz(1976)]{1976ApJS...32..651K}
Kurtz, D.~W. 1976, \apjs, 32, 651

\bibitem[Kurtz(1978)]{1978ApJ...221..869K}
Kurtz, D.~W. 1978, \apj, 221, 869

\bibitem[Kurtz(1980)]{1980MNRAS.193...29K}
Kurtz, D.~W. 1980, \mnras, 193, 29

\bibitem[Kurtz(1989)]{1989MNRAS.238.1077K}
Kurtz, D.~W. 1989, \mnras, 238, 1077

\bibitem[Kurtz et al.(1995)]{1995MNRAS.276..199K}
Kurtz, D.~W., Garrison, R.~F., Koen, C., Hofmann, G.~F., \& Viranna, N.~B.
1995, \mnras, 276, 199

\bibitem[Kurtz \& Martinez(2000)]{2000BaltA...9..253K}
Kurtz, D.~W., \& Martinez, P. 2000, Baltic Astronomy, 9, 253

\bibitem[Landstreet(1998)]{1998A&A...338.1041L}
Landstreet, J.~D. 1998, \aap, 338, 1041

\bibitem[Landstreet et al.(2009)]{2009A&A...503..973L}
Landstreet, J.~D., Kupka, F., Ford, H.~A., et al. 2009, \aap, 503, 973

\bibitem[Lenz \& Breger(2005)]{2005CoAst.146...53L}
Lenz, P., \& Breger, M. 2005, CoAst, 146, 53

\bibitem[Montgomery \& O'Donoghue(1999)]{1999DSSN...13...28M}
Montgomery, M.~H., \& O'Donoghue, D. 1999, DSSN, 13, 28
        
\bibitem[Moon(1985)]{1985CommULO..78}
Moon, T.~T. 1985, Commun. Univ. London Obs., 78

\bibitem[Rufener(1988)]{1999yCat.2169....0R}
Rufener, F. 1998, Observations in the Geneva Photometric System 4.
[CDS Catalog II/169/]

\bibitem[Pollacco et al.(2006)]{2006PASP..118.1407P}
Pollacco, D.~L., Skillen, I., Collier Cameron, A., et al. 2006, \pasp, 118, 1407

\bibitem[Press et al.(1992)]{1992nrfa.book.....P}
Press, W.~H., Teukolsky, S.~A., Vetterling, W.~T., Flannery, B.~P.
1992, Numerical recipes in FORTRAN, 2nd Ed.,
(Cambridge Univ. Press, Cambridge)

\bibitem[Press \& Rybicki(1989)]{1989ApJ...338..277P}
Press, W.~H., \& Rybicki, G.~B. 1989, \apj, 338, 277

\bibitem[Renson \& Manfroid(2009)]{2009A&A...498..961R}
Renson, P., \& Manfroid, J. 2009, \aap, 498, 961

\bibitem[Rodr\'iguez et al.(2000)]{2000A&AS..144..469R}
Rodr\'iguez, E., L\'opez-Gonz\'alez, M.~J., \& L\'opez de Coca, P. 2002, \aaps,
144, 469

\bibitem[Smalley \& Kupka(1997)]{1997A&A...328..349S}
Smalley, B., \& Kupka, F. 1997, \aap, 328, 349

\bibitem[Talon et al.(2006)]{2006ApJ...645..634T}
Talon, S., Richard, O., \& Michaud, G. 2006, \apj, 645, 634

\bibitem[Tamuz et al.(2005)]{2005MNRAS.356.1466T}
Tamuz, O., Mazeh, T., \& Zucker, S. 2005, \mnras, 356, 1466

\bibitem[Torres et al.(2010)]{2010A&ARv..18...67T}
Torres, G., Andersen, J., \& Gim\'enez, A. 2010, \aapr, 18, 67

\bibitem[Turcotte et al.(2000)]{2000A&A...360..603T}
Turcotte, S., Richer, J., Michaud, G., \& Christensen-Dalsgaard, J.
2000, \aap, 360, 603

\bibitem[van Leeuwen(2007)]{2007A&A...474..653V}
van Leeuwen, F. 2007, \aap, 474, 653

\bibitem[Wegner(1981)]{1981ApJ...247..969W}
Wegner, G. 1981, \apj, 247, 969

\end{thebibliography}
\end{document}